\documentclass[twocolumn]{aastex62}
\usepackage{amsmath}
\usepackage{comment}
\usepackage{xcolor}
\usepackage{hyperref}
\usepackage{anyfontsize}

\definecolor{reddish}{rgb}{0.813642,0.212345,0.196948}
\definecolor{blueish}{rgb}{0.0745098,0.52549,0.776471}
		
\hypersetup{
        colorlinks=true,        
        urlcolor=blueish,          
        pdfpagelabels=true,
        hypertexnames=true,
        plainpages=false,
        naturalnames=true,
        pdftitle={K1625 Loose Ends},    
        pdfauthor={Cool Worlds Lab},     
        linkcolor=reddish,          
        citecolor=blueish,        
}

\newcommand{\kepler}{{\it Kepler}}
\newcommand{\gaia}{{\it Gaia}}

\newcommand{\pdf}{\mathrm{Pr}}
\newcommand{\Pmin}{P_{\mathrm{min}}}
\newcommand{\Pmax}{P_{\mathrm{max}}}
\newcommand{\aposteriori}{{\it a posteriori}}
\newcommand{\apriori}{{\it a priori}}

\newcommand{\luna}{{\tt LUNA}}
\newcommand{\multi}{{\sc MultiNest}}
\newcommand{\celerite}{{\tt celerite}}

\newcommand{\specmatch}{{\tt SpecMatch}}

\newcommand{\keplerports}{{\tt KeplerPORTs}}

\shorttitle{Loose Ends for K1625b}
\shortauthors{Teachey et al.}

\begin{document}

\title{Loose Ends for the Exomoon Candidate Host Kepler-1625b}

\correspondingauthor{Alex Teachey}
\email{ateachey@astro.columbia.edu}

\author[0000-0003-2331-5606]{Alex Teachey}
\affil{Department of Astronomy, Columbia University, 550 W 120th Street, New York, NY}

\author[0000-0002-4365-7366]{David Kipping}
\affiliation{Department of Astronomy, Columbia University, 550 W 120th Street, New York, NY}

\author[0000-0002-7754-9486]{Christopher J. Burke}
\affiliation{Kavli Institute for Astrophysics and Space Research, Massachusetts Institute of Technology, Cambridge, MA}

\author[0000-0003-4540-5661]{Ruth Angus}
\affiliation{Department of Astrophysics, American Museum of Natural History, 79th St at Central Park West, New York, NY}
\affiliation{Center for Computational Astrophysics, Flatiron Institute, 162 Fifth Ave., New York, NY}


\author[0000-0001-8638-0320]{Andrew W. Howard}
\affiliation{California Institute of Technology, Pasadena, CA}




\begin{abstract}
The claim of an exomoon candidate in the Kepler-1625b system has generated substantial discussion regarding possible alternative explanations for the purported signal. In this work we examine in detail these possibilities. First, the effect of more flexible trend models is explored and we show that sufficiently flexible models are capable of attenuating the signal, although this is an expected byproduct of invoking such models. We also explore trend models using X \& Y centroid positions and show that there is no data-driven impetus to adopt such models over temporal ones. We quantify the probability that the $500$\,ppm moon-like dip could be caused by a Neptune-sized transiting planet to be $<0.75$\%. We show that neither autocorrelation, Gaussian processes nor a Lomb-Scargle periodogram are able to recover a stellar rotation period, demonstrating that K1625 is a quiet star with periodic behavior $<200$\,ppm. Through injection and recovery tests, we find that the star does not exhibit a tendency to introduce false-positive dip-like features above that of pure Gaussian noise. Finally, we address a recent re-analysis by \citet{kreidberg} and show that the difference in conclusions is not from differing systematics models but rather the reduction itself. We show that their reduction exhibits i) slightly higher intra-orbit and post-fit residual scatter ii) $\simeq900$\,ppm larger flux offset at the visit change iii) $\simeq2$ times larger Y-centroid variations and iv) $\simeq3.5$ times stronger flux-centroid correlation coefficient than the original analysis. These points could be explained by larger systematics in their reduction, potentially impacting their conclusions. 
\end{abstract}


\section{Introduction} 
\label{sec:intro}
Last year, \citet{TK18} (TK18 hereafter) presented evidence for a large exomoon
orbiting the gas giant Kepler-1625b. That work was based on a joint analysis of
three transits of the planet observed with \kepler, and a fourth transit observed
with the Hubble Space Telescope (HST) in October 2017 (GO-15149, PI Teachey). 
The conclusion was based on the presence of significant transiting timing variations (TTVs)
in the system, as well as a sustained dip in the star's brightness following planetary
egress. These two lines of evidence were interpreted as self-consistent indications
that a large moon is present in the system. A number of alternative explanations
for these two signals were explored and the likelihoods of these alternatives were considered.
Taken together, the exomoon hypothesis emerged as the best explanation for the data in hand.

Since the publication of TK18, discussions with and amongst colleagues have highlighted
open questions and unresolved issues emerging from the analysis. In this work we
take the opportunity to address some of these points and present an update
on the prospects of confirming or rejecting the exomoon hypothesis for Kepler-1625b.

This paper is structured as follows. In Section~\ref{sec:detrending}, we
explore other systematic models to account for the long-term trend seen in
the TK18 light curve and the effects they have on the interpretation. 
In Section~\ref{sec:kreidberg} we address differences between our work and 
that of another group \citep{kreidberg}, whose independent reduction and analysis we became 
aware of during the course of writing this paper. In Section~\ref{sec:planetc}, 
we discuss the possibility that the moon-like dip is in fact caused by a second, 
previously undetected transiting planet in the system. In Section~\ref{sec:stellar}, 
we provide a more detailed assessment of the host star's activity and investigate 
the possibility that it could be responsible for the moon-like dip. In
Section~\ref{sec:followup}, we use forward propagation of the TK18 solution to
determine the location and probability of seeing exomoon transits in
future epochs. Conclusions are summarized in Section~\ref{sec:conclusions}.

\section{Other Systematic Models} 
\label{sec:detrending}
\subsection{Overview}
\label{sub:detrendingoverview}

TK18 employed three different models to account for the long-term trend
seen in their data. These were broadly motivated to follow as closely as
possible the most standard approaches in the literature for previous WFC3
analyses (see \citet{wakeford:2016} for an overview of WFC3 systematics). 
Most authors have previously elected to use a simple linear trend
for this correction, of the form $a_0 + a_t (t-t_0)$ (e.g. see
\citealt{huitson:2013,ranjan:2014,knutson:2014}). In some rarer cases, a
quadratic model has been invoked \citep{stevenson:2014a,stevenson:2014b}
and thus both of these models were attempted. A third exponential model
was also attempted giving three trend models in total, all with time
as the independent variable.

Gaussian processes (GPs) have also been utilized in, for example, 
\citet{evans:2018} to handle WFC3 systematics. In general, however, 
GPs are not obviously appropriate for the moon search unless there is reason to 
suspect the data are not drawn from a sequence of independent Gaussians. 
As we show later, we see no evidence of time-correlated noise structure.
But the flexibility of GPs mean they will inevitably fit out a moon-like dip,
and insomuch as less flexible detrending models explored here are also capable 
of attenuating or removing the moon signal, invoking GPs here is neither 
well-motivated nor particularly illuminating.

TK18 argued that it was crucial to perform this detrending simultaneous
to the transit fits, repeating for each model (planet/moon/TTV), to
account for the fact that the trend model appeared highly covariant with
the moon-like dip. In comparing these models, we re-emphasize here that
full Bayesian evidences should be used. As a non-linear model \citep{luna:2011},
the number of degrees of freedom cannot be estimated and thus reduced $\chi^2$ 
comparisons are certainly invalid \citep{rene:2010}. Another popular
alternative to computing evidences is the Bayesian Information Criterion
\citep[BIC,][]{BIC}. This was used, for example, in \citet{kreidberg}.
But again here there are serious concerns about its use for
this problem. By invoking a Laplacian approximation on the posterior,
one approximates the posterior to a Gaussian centered on the maximum
likelihood estimator, which is inappropriate for highly multi-modal
posteriors such as those resulting from exomoon fits \citep{HEK1}. Further,
the BIC is not guaranteed to yield a Bayes factor which is close to one 
computed using priors an observer would consider appropriate, since it assumes
the unit information prior on the model parameters \citep{weakliem:1999}.
For these reasons, model comparisons are performed using the Bayesian evidence
in what follows.

The three trend models considered by TK18 allowed for an offset
between the two visits, which was most apparent in the extreme channels
and to a lesser degree in the white light curve. Clearly the models explored
by TK18 are a small subset of an essentially infinite number
of possible models one could try. In general, the more flexible the model,
the easier it is to fit out the moon-like dip
when assuming no-moon present. It was for this reason that more flexible
trend models were not explored by TK18, since any sufficiently
flexible instrument model can fit-out interesting astrophysics.

Nevertheless, this was neither demonstrated nor investigated in detail
in that work. For this reason, we re-visit the trend modeling here
exploring a) the effect of going to higher-order polynomials
b) the effect of allowing for discontinuous trend models c) the effect
of changing the dependent variable.

\subsection{Higher-order polynomials}
\label{sub:cubic}

Although an infinite number of polynomials exist beyond a quadratic
trend, we here perform a cubic model as a simple extension to illustrate
the effects. We re-ran the moon ($M$) and zero-radius moon
($Z$) models on the TK18 using \multi\
\citep{feroz:2008,feroz:2009} and \luna\ \citep{luna:2011}, as
was done in TK18, except we add an additional cubic term to
the quadratic trend model.

\begin{figure*}
\begin{center}
\includegraphics[width=2\columnwidth,angle=0,clip=true]{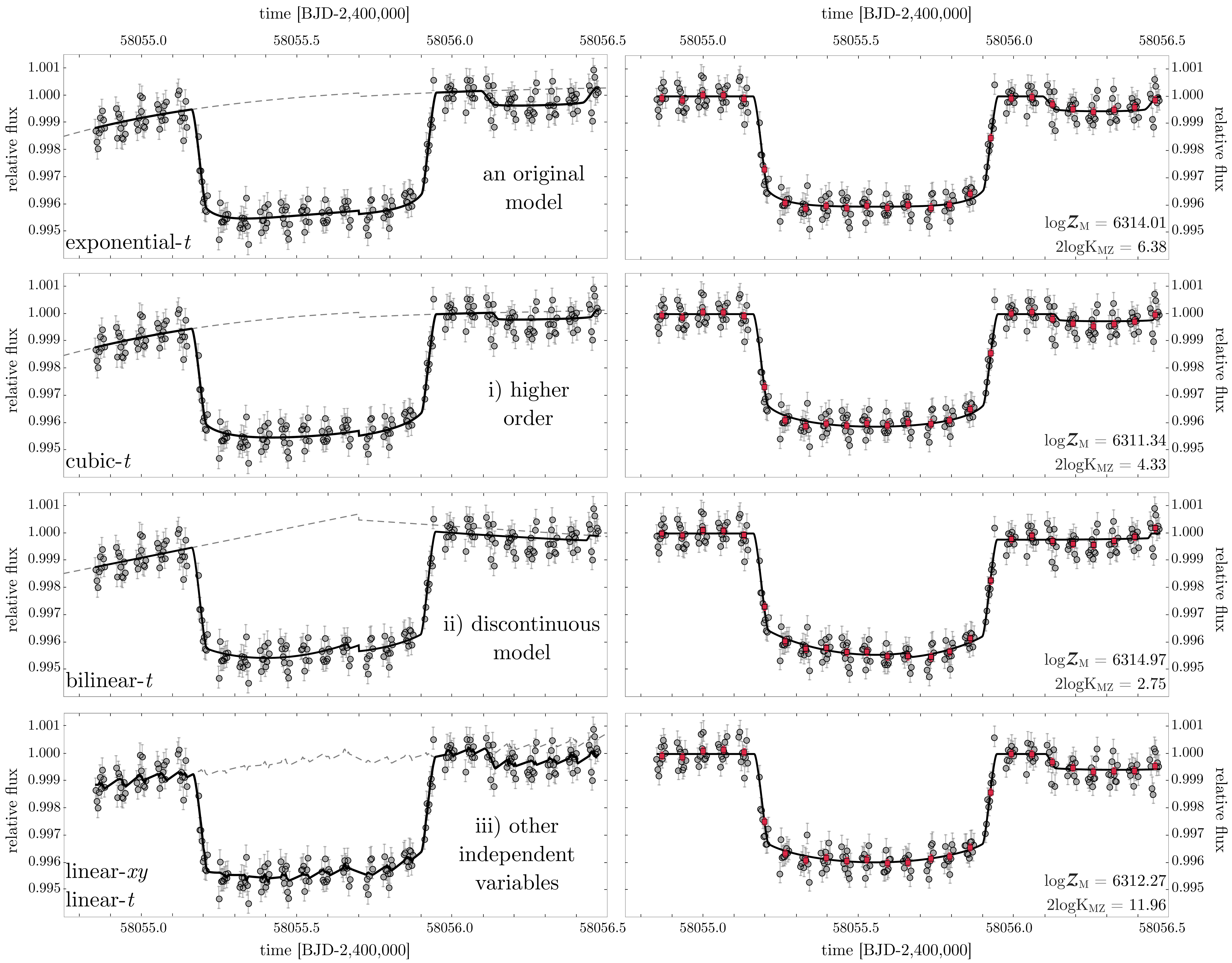}
\caption{\emph{
Comparison of three different long-term trend models applied to the
TK18 data. The left column shows the uncorrected data with the
trend model overlaid while the right column shows the post-correction data.
More flexible systematic models than those considered by TK18
attenuate the formal exomoon evidence and also find alternative
modes that are inconsistent with the TK18 candidate signal.}}
\label{fig:polynomials}
\end{center}
\end{figure*}

The resulting maximum \aposteriori\ light curve is shown in the
second row of Figure~\ref{fig:polynomials}. The shape closely
matches that of the exponential model shown above it (and indeed
the quadratic model form TK18), reflecting the fact that the
cubic term is almost zero. In fact, we can quantify this statement
by evaluating the 1\,$\sigma$ credible interval of the marginal
posterior of the cubic coefficient to be $630_{-450}^{+430}$\,ppm,
reflecting how the posterior is only 1.4\,$\sigma$ from zero.

It is therefore not surprising that a) the cubic fit returns a
similar Bayes factor in favor of the exomoon to the original
models of TK18, and b) the cubic fit has an overall lower evidence
than the other models ($\log \mathcal{Z}_M = 6311.34\pm0.16$) since
it includes effectively wasteful parameter volume. In other words,
the model has been penalized for additional complexity.

\subsection{Discontinuous polynomials}
\label{sub:doublelinear}

TK18 only considered models where the function is continuous across
the visit change, except for a flux offset. This means that the higher-order
polynomial coefficients, $a_{t}$ and $a_{t2}$ in the case of a quadratic model,
are the same on both sides of the visit change. The motivation for this was that
a) the small trends either side of the visit change did not correlate with
centroid position and thus did not appear to be instrumental in nature, and
b) the star's intrinsic variability should not change dramatically on either side
of the visit change.

We discuss here the effect of relaxing this assumption. As in the case of
higher-order polynomials, this essentially represents a more flexible model.
The simplest discontinuous polynomial is two independent straight lines
(``bilinear'' in what follows). As in TK18, all detrending choices
explored in this work are implemented after the hook correction has been applied.
Because the hook correction minimizes intra-orbit RMS independent of the model
under consideration, there is no covariance between them and therefore
the evidences are not impacted. We refer readers to TK18 for a more thorough
discussion of this choice.

We find that the fits favor a very pronounced reversal in the systematic
gradient located at the visit-change, as can be seen from the third row
of Figure~\ref{fig:polynomials}. It is unclear how this behavior could
manifest physically, since flux is apparently uncorrelated with centroids
in each individual visit for the comparison star KIC\,4760469 (see
Figure~S10 and Section 1.2.10 of \citealt{TK18}). That is, while there is no
known explanation for the visit-long trends, we would \apriori\ expect the
target star and the comparison star to display similar systematic 
morphologies. The fact that they do not leads us to question whether 
a downturn in the target star trend can be attributed to instrumental 
systematics. At the same time, we see no reason to expect the
star to exhibit a pronounced reversal coincidental with HST's visit
change. The second visit slope has a negative gradient that absorbs the decrease
in brightness caused by the moon-like dip, and for this reason the evidence
is significantly attenuated for the exomoon model. Indeed, the moon solution
is quite distinct from the original paper and can be immediately dismissed
as suspicious because the moon ingress is nearly coincident with the flux
offset associated with the visit change.

The bilinear model has two free parameters per visit,
giving four in total - the same as the number of free parameters describing
the TK18 quadratic and exponential models. Despite having the same
number of free parameters, it does not necessarily have the same degrees of
freedom. These two concepts are distinct if the underlying model is non-linear
\citep{rene:2010}, which is true here because of the non-linear step function
occurring at the visit change. In fact, it can be seen that the bilinear
model in fact has more freedom, because it does not require a continuous
gradient across the boundary, unlike the case of the TK18 quadratic
model.

It is therefore perhaps not surprising that this more flexible model is
able to fit-out the moon-like dip sufficiently well to find no evidence for
the putative moon. This analysis does not particularly add to or detract
from the exomoon hypothesis, since the behavior can be understood as a
byproduct of employing more flexible trend models. Clearly the attenuation
of the moon-like dip does not necessarily imply that the trend model is 
incorrect; strictly speaking the conclusion is simply that the moon+flexible
model is not supported by the data. Put another way, the moon model and
the planet-only model are essentially equiprobable with this trend model,
and therefore we would not claim evidence for a moon. We find no physical
or data-driven motivation for adopting the bilinear model, nor is 
there precedent for doing so in the literature. On the other hand,
the observation is unprecedented in several ways (e.g. the faintness of the target, 
the 40 hour duration, and the objective itself), so we
cannot rule out the possibility that we are observing unique or heretofore
only marginally important systematic effects.

\subsection{Changing the independent variable}
\label{sub:xyfits}
We now consider a third and final modification to the systematic model,
namely, modeling the systematics as a function of both time and centroid
position, rather than simply a function of time. We start by taking our 
simplest model, the linear-$t$ model, given by

\begin{align}
S(t) = a_0 + a_{t1} (t-t_0),
\end{align}

and extending it to include a linear dependency on $x$ and $y$ centroid
positions:

\begin{align}
S(t,x,y) =& a_0 + a_{t1} (t-t_0) \nonumber\\
\qquad& + a_{x1} (x-x_0) + a_{y1} (y-y_0),
\end{align}

where $a_i$ are coefficients to fit for and the subscript 0 variables
represent the median time/centroid positions. Using the same photodynamical
\multi\ fitting software from TK18, the resulting maximum \aposteriori\
light curve is shown the fourth row of Figure~\ref{fig:polynomials}.

The figure, as well as the evidences quoted in the panel, show that the same
moon is again detected. We therefore conclude that adding $x$ and $y$ as
linear independent variables to the systematic model does not significantly
affect the conclusions of TK18.

\subsection{Fixing orbits to coplanar}
\label{sub:coplanar}

The inclined solution for the exomoon candidate K1625b-i is particularly
curious. The fact the posteriors favor an inclined solution suggests that
it should be very difficult to fit the same moon to the existing data
(both \kepler\ and HST) if one imposes coplanarity. To investigate this,
we repeat the three trend model fits of TK18 for the M models but fix
$i=90^{\circ}$. Comparing the resulting evidences to the original Z models
of that work ($2 \log \mathcal{Z}_{MZ} = 17.77 \pm 0.33$, $3.61 \pm 0.33$, and $6.38 \pm 0.34$
for the linear, quadratic and exponential models, respectively) indeed show the case for an exomoon is removed (see
Table~\ref{tab:coplanar}). This highlights the importance of including
inclination in such fits.

\begin{table}
\caption{\emph{Repeat of the TK18 model fits but enforcing the condition that
the moon must be coplanar i.e. $i_S=90^{\circ}$.}} 
\centering 
\begin{tabular}{c c} 
\hline\hline 
reduction & $2\log(\mathcal{Z}_{M,\mathrm{coplanar}} - \mathcal{Z}_Z)$  \\ [0.5ex] 
\hline 
linear-$t$		& $1.05 \pm 0.32$ \\
quadratic-$t$	& $-0.28 \pm 0.33$ \\
exponential-$t$	& $-0.01 \pm 0.33$ \\ [1ex]
\hline\hline 
\end{tabular}
\label{tab:coplanar} 
\end{table}

\subsection{Using the comparison star as model benchmark}
\label{sub:comparisonstarmodels}

Since the comparison star is expected to be stable (TK18),
it provides a useful control test for comparing the different
possible systematic trend models. Expanding to quadratic order
in $x$, $y$ and $t$, we fitted nine different models to the
comparison star (assuming an intrinsically flat baseline)
using \multi. The various models and resulting evidences are listed in
Table~\ref{tab:comp}.

\begin{table*}
\caption{\emph{Bayesian evidences from applying various systematic models to the comparison star
KIC 4760469. All evidences are quoted with $1334.18$ subtracted - the absolute value obtained
for the second model listed - ``quadratic-$t$''.
Models with a $^{*}$ indiciate that this one of the original models
used by TK18. Models with a $^{\dagger}$ are those used by \citet{kreidberg}. The model
considered by \citet{kreidberg} is formally indistinguishable from the systematic models used in TK18, and
are therefore not favored over those used in that work.}} 
\centering 
\begin{tabular}{l l c} 
\hline\hline 
label & systematics model & $\log\mathcal{Z}$ \\ [0.5ex] 
\hline 
linear-$t$$^{*}$		& $a_0 + a_{t1} (t-t_0) + (a_0'-a_0) \mathcal{H}[t-t_{II}]$					   & $-0.60 \pm 0.06$ \\ 
quadratic-$t$$^{*}$	& $a_0 + a_{t1} (t-t_0) + a_{t2} (t-t_0)^2 + (a_0'-a_0) \mathcal{H}[t-t_{II}]$ & $0.00 \pm 0.06$ \\ 
exponential-$t$$^{*}$	& $a_0 + a_{e1} \exp(\tfrac{t-t_0}{a_{e2}}) + (a_0'-a_0) \mathcal{H}[t-t_{II}]$& $-0.38 \pm 0.06$ \\ 
\hline
linear-$xy$$^{\dagger}$					& $a_0 + a_{x1} (x-x_0) + a_{y1} (y-y_0)$    								 & $-0.61 \pm 0.06$ \\ 
linear-$xy$ linear-$t$		& $a_0 + a_{t1} (t-t_0) + a_{x1} (x-x_0) + a_{y1} (y-y_0)$				     & $-4.29 \pm 0.06$ \\ 
linear-$xy$ quadratic-$t$	& $a_0 + a_{t1} (t-t_0) + a_{t2} (t-t_0)^2 + a_{x1} (x-x_0) + a_{y1} (y-y_0)$& $-3.96 \pm 0.07$ \\ 
\hline
quadratic-$xy$				& $a_0 + a_{x1} (x-x_0) + a_{x2} (x-x_0)^2 + a_{y1} (y-y_0) + a_{y1} (y-y_0)^2$    								 & $-1.27 \pm 0.06$ \\ 
quadratic-$xy$ linear-$t$	&$a_0 + a_{t1} (t-t_0) + a_{x1} (x-x_0) + a_{x2} (x-x_0)^2 + a_{y1} (y-y_0) + a_{y1} (y-y_0)^2$					 & $-4.97 \pm 0.07$ \\ 
quadratic-$xy$ quadratic-$t$&$a_0 + a_{t1} (t-t_0) + a_{t2} (t-t_0)^2 + a_{x1} (x-x_0) + a_{x2} (x-x_0)^2 + a_{y1} (y-y_0) + a_{y1} (y-y_0)^2$& $-3.43 \pm 0.07$ \\ [1ex] 
\hline\hline 
\end{tabular}
\label{tab:comp} 
\end{table*}

In TK18, only three of these models were considered but
it turns out none of the other six models proposed here yield
an evidence superior to the simple time models. We conclude that
analysis of the comparison star indicates that
systematic models using $x$ and $y$ centroid positions are not
supported by the current data.

\section{Comparison to Kreidberg et al. (2019)} 
\label{sec:kreidberg}
During the final preparations of this paper, it came to our attention that
\citet{kreidberg} (henceforth KLB19) had conducted an independent reduction of the HST WFC3
observations of Kepler-1625 and concluded that there was no evidence for an
exomoon based on the apparent lack of a moon-like dip following planetary egress. 
We will compare the KLB19 reduction and results to that of TK18 in what follows.

\subsection{Raw photometry comparison}
\label{sub:photocomp}

\begin{figure*}
\begin{center}
\includegraphics[width=2\columnwidth,angle=0,clip=true]{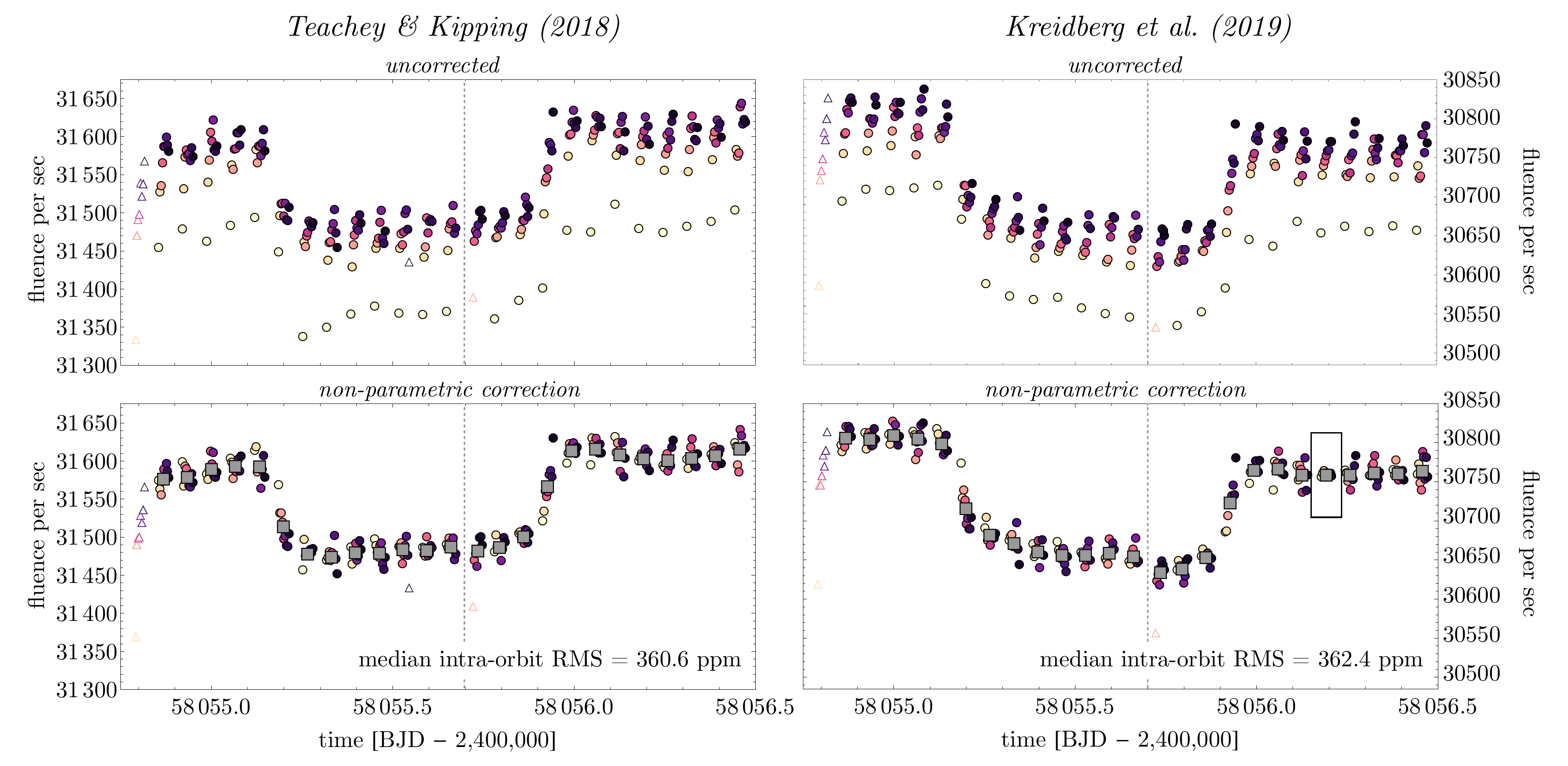}
\caption{\emph{A comparison of light curves before and after the hook correction
from TK18 and KLB19. Compare to Figure 2 in TK18. As in that work,
the data are color coded by the observation number within each HST orbit 
(light yellow for the first observation, dark purple for the last).
The grey squares in the bottom panel represent the binned flux for each orbit. 
Triangles indicate observations from the first orbit, which are left out of
the hook correction normalization. the anomalously low scatter in the 22nd
orbit of the KLB19 analysis is highlighted with a rectangle.}}
\label{fig:photocomp}
\end{center}
\end{figure*}

It is instructive to make a side-by-side comparison of the raw photometry
presented in TK18 and the new reduction by KLB19
before any hook or trend corrections have been applied, which
is shown in the top panel Figure~\ref{fig:photocomp}.

Before any systematic effects have been corrected, the photometry
from both groups is apparently quite similar, yet important differences are evident. Quick inspection
of both light curves reveals a much more pronounced offset in flux for KLB19 
occurring at the instant of the visit change between orbits 14 and 15 (marked by the vertical dashed lines in
that figure). As described in TK18, the full guide star acquisition
performed at the beginning of orbit 15 was responsible for the introduction
of this offset. Detailed modeling (described in Section~\ref{sub:fits}) 
finds that the amplitude of the offset increase in every case from 
the TK18 reduction to that found in KLB19:

\begin{itemize}
\item from $(20\pm110)$\,ppm in TK18 to $(-900\pm120)$\,ppm in KLB19, for the linear-$t$ Z model; 
\item from $(-140\pm120)$\,ppm in TK18 to $(-850\pm130)$\,ppm in KLB19, for the quadratic-$t$ Z model; and
\item from $(-20\pm110)$\,ppm in TK18 to $(-880\pm110)$\,ppm in KLB19, for the exponential-$t$ Z model.
\end{itemize}


There is of course only one ground-truth in terms of the motion of the
telescope, and astrophysical variation. Because we can reasonably assume the 
star itself is not exhibiting a sudden change in flux after the 14\textsuperscript{th} HST orbit, the 
discontinuity there must be systematic. Thus, a larger discontinuity could be viewed
as being farther from the star's ground-truth, requiring a more substantial correction
that could impact the results of KLB19.

\subsection{Hook correction} 
\label{sub:hookcomp}

We next applied the exact same hook correction algorithm described in TK18
to the KLB19 reduction. KLB19 also uses the non-parametric approach of TK18, 
thereby providing a fair comparison of the two, and this is shown in the 
bottom panel of Figure~\ref{fig:photocomp}).

The mean intra-orbit photometric RMS from KLB19 is somewhat
smaller at 360.7\,ppm, versus 374.8\,ppm for TK18\footnote{TK18 quote 375.5\,ppm 
but that value is the mean intra-orbit RMS after 10 rounds of hook correction 
iterations, whereas the final light curve actually uses 100 rounds.}. At first glance, 
this appears to indicate that the KLB19 reduction is more precise. However, 
inspection of Figure~\ref{fig:photocomp} reveals that the 22nd HST orbit appears to 
display an anomalously low scatter of just 85\,ppm. While TK18 also find that this 
orbit has the lowest scatter, the RMS is much more consistent with the other orbits, 
at 210\,ppm.

For normally distributed data, the standard deviation of sample standard deviations 
equals $\sigma/\sqrt{2(n-1)}$. 
Since the mean RMS for TK18 is $\sigma=374.8$\,ppm, and the average number of points 
per orbit is 8.8, then one should expect RMS values with a standard deviation of $94.9$\,ppm. 
The actual standard deviation of RMS values is less than one percent larger at $95.5$\,ppm. 
his in turn means that the 210\,ppm smallest RMS value is only 1.7\,$\sigma$ from the mean.

For the KLB19 orbits, the expected standard deviation in RMS values is 
91.3\,ppm and the observed value is 8.6\% higher at 99.2\,ppm. Critically, the 22nd 
orbit now appears to be a 3\,$\sigma$ outlier. Strictly speaking, the formula above 
is only valid for $n \gg 1$, so we are at the limit of applicability in the present case. 
Thus, a better estimate for how surprising this 
orbit is can be obtained by masking the orbit, taking the mean of the other RMS 
values, and then generating fake Gaussian data for all orbits and Monte Carlo evaluating 
the expected distribution of RMS values. This reveals that the 22nd 
orbit from KLB19 is anomalously low at the 4.0\,$\sigma$ level.

This seems highly implausible from a statistical perspective and would make the 
22nd orbit intra-orbit RMS an outlier by most definitions. In the presence of outliers, 
a more robust summary statistic to compare the precision of each reduction is the 
\textit{median} intra-orbit RMS, rather than the mean. On this basis, the original 
TK18 reduction is marginally more precise at 360.6\,ppm versus 362.4\,ppm for KLB19.
That is, they are essentially indistinguishable from the perspective of their noise profile.
The source of this improbably low scatter in the 22\textsuperscript{nd} orbit is unknown, but it appears to be present 
in the data before the application of the hook correction. 

\subsection{Centroids} 
\label{sub:centcomp}

\begin{figure*}
\begin{center}
\includegraphics[width=2\columnwidth,angle=0,clip=true]{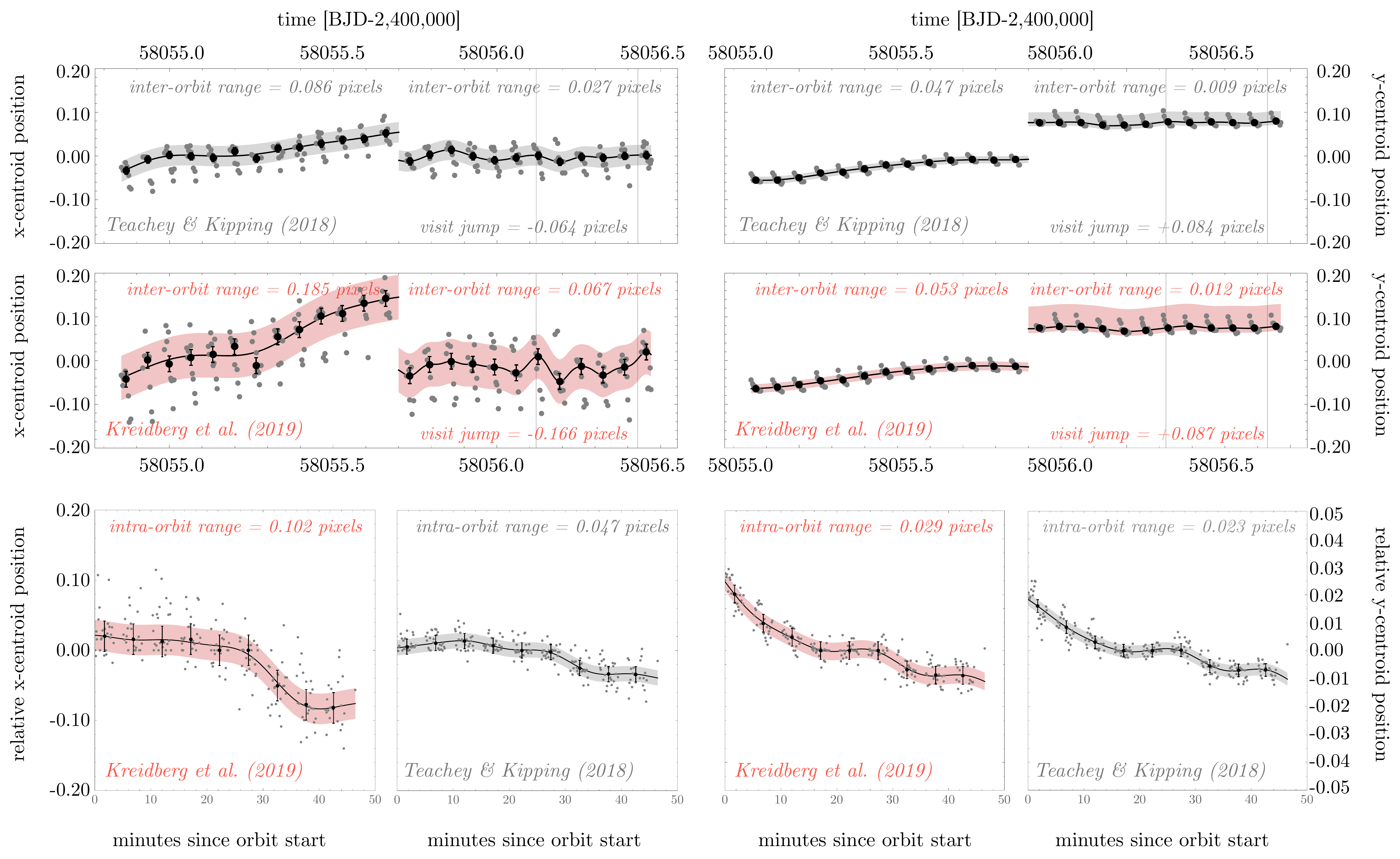}
\caption{\emph{
Comparison of the centroids reported by TK18 (gray) and those
of KLB19 (red). In all cases, we find that the KLB19
centroids exhibit larger variations.
}}
\label{fig:centroids}
\end{center}
\end{figure*}

Centroids deserve special attention since KLB19 
use the target's position on the detector as the basis for their
systematics model correction. TK18 presented their centroid variability 
in Figure~S10 of that work, for both Kepler-1625 and the comparison star
KIC\,4760469. Figure~\ref{fig:centroids} directly compares the
centroids of TK18 to those of KLB19 for Kepler-1625, where 
morphological similarities are apparent.

As was found in TK18, there is substantial variation of the
apparent centroid position within an orbit, which we attribute to
the hook and / or breathing effects rather than a real variation. For this
reason, long-term behavior (associated with pointing drift) is best tracked 
using the orbit median centroid positions, shown in black in
Figure~\ref{fig:centroids}.

We find that the range in inter-orbit $Y$-centroid position is 10\% higher
for the first visit and 30\% higher for the second visit in the KLB19 reduction than 
that of TK18, and 2.2 to 2.5 times higher for the $X$-centroid position. 
Similarly, the change in centroid position after the visit change is 
2.6 times higher in $X$ for KLB19 than TK18.

We tried offsetting the median centroid of each orbit and then
orbit-folding (see lowest row of Figure~\ref{fig:centroids})
to look at intra-orbit centroid variations, rather than inter-orbit.
As before, we find higher intra-orbit centroid variability for the
KLB19 reduction, increased by similar levels.

The origin of these centroid discrepancies is unclear. Systematic effects
such as the hook and breathing effects likely play a role in the
calculated position for each image, as suggested by the intra-orbit 
centroid variations which do not appear to be associated with pointing drift. 
Different handling of these systematics could therefore reasonably explain the discrepancy.
due to the image rotation peformed in TK18, it is not possible
to apply our centroid corrections to the KLB19 reduction, nor is it
possible to use the KLB19 centroids for detrending our extracted light curve.

We also point out that the calculation of these centroids is
handled differently in TK18 and KLB19. TK18 simply calculates
the flux weighted centroid of the optimal aperture at every time step as

\begin{equation}
X_{centroid} = \frac{\sum_{i}^{N} x_i f_i}{\sum_{i}^{N} f_i}
\end{equation}

\noindent where $x_i$ a an $f_i$ are the pixel coordinate and flux of pixel $i$, 
respectively. Calculation of the centroid in $Y$ is identical.
TK18 performed this operation for both the target star and the comparison star
KIC4760469, which showed good agreement. 

By contrast, KLB19 perform a more complicated analysis to compute the motion, 
with different methodologies for the $x$ and $y$ directions. For the $y$ (or ``spatial'') 
direction, KLB19 sum the flux in each column of the image at each time step, 
perform a 4-pixel Gaussian convolution of the resulting array and then an interpolation to 
compute a best fitting offset from a template at each time step. It is not 
obvious how spatial information is recovered from this algorithm as described in
KLB19, nor whether comparison to a template could introduce biases. KLB19 perform a 
similar operation for the $x$ (or ``spectral'') direction, though now only summing
up along the target spectrum instead of along each row in the image.
The result, as shown in Figure~\ref{fig:centroids} is morphologically similar to 
TK18, though with larger systematics. 

\subsection{Systematic trend comparison} 

KLB19 use systematic models which decorrelate against $X$ and $Y$
centroid position rather than just time. Since the previous subsection has
conducted a like-for-like comparison of this model, it is instructive to inspect
the systematic parameter posteriors that result. We list these values in
Table~\ref{tab:xygradients}.

\begin{table}
\caption{\emph{Using the systematic model ``linear-$xy$'' only (which assumes no part
of the trend is dependent on time), we compare here
the parameters $a_{x1}$ and $a_{y1}$ (i.e. the centroid gradient terms) which
result from three different reductions. Elements list median and 68.3\% credible
intervals in units of parts per million per day.}} 
\centering 
\begin{tabular}{c c c} 
\hline\hline 
reduction & $a_{x1}$ & $a_{y1}$ \\ [0.5ex] 
\hline 
TK18, Kepler-1625				& $+2700_{-900}^{+910}$ &  $+4780_{-450}^{+440}$ \\
TK18, KIC 4760469				&  $+580_{-960}^{+970}$ &  $+3050_{-620}^{+610}$ \\
KLB19, Kepler-1625	            &  $+830_{-400}^{+400}$ & $-10170_{-410}^{+410}$ \\ [1ex]
\hline\hline 
\end{tabular}
\label{tab:xygradients} 
\end{table}

As can seen be seen from the table, KLB19 find an overall stronger
dependency between flux and centroid position than TK18, with almost all of the
variability coming in via the $Y$-direction. It is also worth noting that the
sign of $a_{y1}$ reverses for the KLB19 reduction.

We also remind the reader that our earlier comparison of different trend models
applied to the comparison star found that models including $X$-$Y$ pixel
position were disfavored (Section~\ref{sub:comparisonstarmodels}) over temporal
models. In any case, it is difficult to determine the degree to which these
discrepancies arise from the different centroiding approach, and how much 
is due to differences in the raw fluxes owing to the reduction itself.

\subsection{Model evidences comparison}  
\label{sub:fits}

TK18 perform model comparison using the Bayes factor calculated using
Bayesian evidences (marginal likelihoods). In contrast, KLB19 perform their
model selection using reduced $\chi^2$ and the Bayesian Information
Criterion (BIC). As discussed earlier in Section~\ref{sub:detrendingoverview},
model comparison using the reduced $\chi^2$ is invalid for non-linear
models and it is thus not appropriate for transit light curve fits. The
BIC is also inappropriate due to the multimodality of the posterior, which
is poorly described by the Laplacian approximation used by BIC. Further, it
is generally not guaranteed to produce an approximation of the Bayes factor
\citep{weakliem:1999}, and indeed it has been argued to not even
represent an approximation to any exact Bayesian solution - including the
Bayes factor \citep{gelman:1995}. Accordingly, we strongly urge the avoidance
of these tools for exomoon model selection.

To perform a full comparison between the two reductions, it is instructive to
repeat the full photodynamical \multi\ fits conducted by TK18 on the
KLB19 reduction. This allows us to evaluate what the Bayes
factor would be for the exomoon had we used this data set instead. 

We fit the hook-corrected light curves of KLB19 using the same three
models used by TK18 - linear in time, quadratic in time and exponential in time,
all of which also include a flux offset parameter at the visit change.
Further, we ran the photodynamical \multi\ fits for both the TK18 and
KLB19 adopting a fourth systematics model - one motivated
by the choice of KLB19 to decorrelate against centroid position.
Specifically, this model is linear in time as well as in $X$ and $Y$ centroid position
i.e. an example of changing the independent variable. The results of these fits
are summarized in Table~\ref{tab:fits}.

\begin{table}
\caption{\emph{Top: Bayesian model evidences using different formalisms
for the systematic model, and comparing two different reductions of the
HST WFC observations of K1625. Each element represents
$2\log(\mathcal{Z}_M - \mathcal{Z}_Z)$ - the Bayes factor
for the exomoon. $^{*}$ = original fits from TK18.
}} 
\centering 
\begin{tabular}{c c c} 
\hline\hline 
model & TK18 & \citet{kreidberg} \\ [0.5ex] 
\hline 
linear-$t$$^{*}$		& $17.77\pm0.33$ & $1.08\pm0.32$ \\
quadratic-$t$$^{*}$		&  $3.61\pm0.33$ & $1.38\pm0.32$ \\
exponential-$t$$^{*}$	&  $6.38\pm0.34$ & $1.88\pm0.33$ \\

linear-$xy$ linear-$t$	& $11.96\pm0.34$ & $0.56\pm0.34$ \\[1ex]

\hline\hline 
\end{tabular}
\label{tab:fits} 
\end{table}

As can be seen from the table, the KLB19 reduction consistently
yields lower Bayes factors for the moon solution versus that found by TK18.
Although a moon-dip is favored in all cases (contrasting with the BIC and
reduced $\chi^2$ testing of KLB19), the strength of the evidence
is diminished to such a degree that we would not consider it justifiable to
claim evidence for an exomoon. Combined with the investigation described earlier
in Section~\ref{sub:xyfits}, this strongly suggests that the differing conclusions
between TK18 and KLB19 is not due to the choice of systematic model,
but rather due to the reduction itself. This is the same conclusion reached
by KLB19. However, we do not agree that the moon's existence has been ruled out, 
particularly in light of a second independent reduction and analysis carried out by \citet{heller:2019},
which also finds evidence for the moon-like dip following the planet's transit.

We point out that \citet{nelson:2018} found through an extensive comparison
of approaches to computing model evidences that uncertainties are likely to be
underestimated. As such the uncertainties quoted in Tables \ref{tab:comp} and
\ref{tab:fits} may be too low. For each run we used 4000 live points, which 
is twice the recommended number for accurate evidence uncertainties \citep{feroz:2008}. In any case, 
artificially low uncertainties would not invalidate the salient
features of our argument here, namely, that 1) we see no strong impetus to 
adopt a detrending model based on centroids (see Section \ref{sub:centcomp}), and 
2) evidence for the moon is considerably weaker based on the KLB19 light curve.

\subsection{Model residuals comparison}   
\label{sub:fits}

The null hypothesis is that no moon is present around K1625 and so the
obvious place to conduct a residual analysis is on the no-moon models (model Z).

The original residual analysis conducted by TK18 (see Figure~S17) shows that
without a moon there appears to be high time-correlated noise when inspecting
simple RMS vs bin-size style diagrams. However, as shown in that same figure,
the origin of the time-correlated noise excess is apparently localized in time
to the specific point where TK18 claim evidence for a moon-like dip.

A fairer test of residual noise is then to continue using the null hypothesis
but mask out the region where TK18 claim a photometric anomaly associated with a
possible moon. To accomplish this, we compute the maximum \aposteriori\ model
residuals for the exponential-$t$ model (since this is the model used for light
curve comparison by KLB19 in their Figures~3 \& 4), for both the TK18
and KLB19 reductions, and then mask out the region $t>2458056.1$\,BJD.
This also conveniently removes orbit 22 of KLB19, which is argued to
show anomalously low scatter in Section~\ref{sub:hookcomp}.

\begin{figure}
\begin{center}
\includegraphics[width=\columnwidth,angle=0,clip=true]{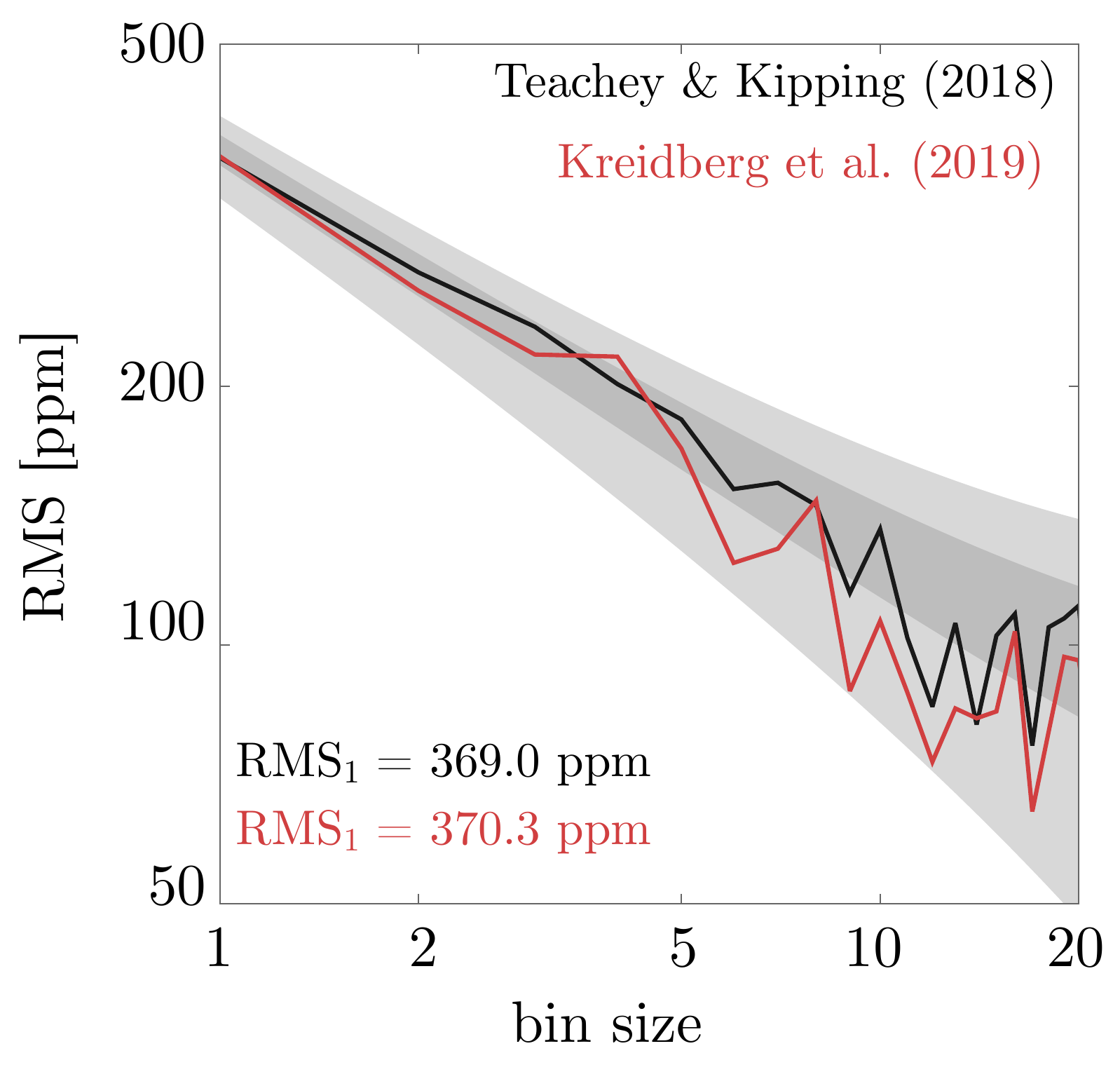}
\caption{\emph{
RMS vs bin-size diagrams for the exponential-$t$ model regressed to the
TK18 reduction (black) and the KLB19 reduction (red). In both
cases the model assumes no moon but masks the region $t>2458056.1$\,BJD
where the moon-like dip is seen by TK18. Both reductions appear consistent
with Gaussian noise properties (the gray 1 and 2\,$\sigma$ regions shown).
}}
\label{fig:residuals}
\end{center}
\end{figure}

We then compute an RMS vs bin-size diagram, as shown in Figure~\ref{fig:residuals}.
We find that both reductions display Gaussian-like behavior with no clear
indications of excess noise. Without any binning, the RMS values are
369.0\,ppm for TK18 and 370.3\,ppm for KLB19, which are effectively
identical.

\subsection{Presence of TTVs}  

The sustained moon-like dip in the HST observation observed by TK18 
is one important element of the case for the exomoon. However, 
another critical, self-consistent component of the case for the exomoon 
presented in TK18 is the presence of TTVs in the system. A large 
moon like the one described in TK18 is expected to exert a significant
gravitational influence on the planet, detectable in the photometry 
in the form of timing variations.

As described in TK18, the transit of Kepler-1625b in the HST observation
occurred a full 78 minutes earlier than anticipated based on a linear
ephemeris calculated from the three transits of the planet observed by \kepler. 
This indicates the presence of TTVs to $\sim 3 \sigma$ confidence.

We fit the transit timings for the KLB19 light curve and find 
$\tau = 58055.5539^{+0.0013}_{-0.0012}$, 
$58055.5538^{+0.0013}_{-0.0012}$, and
$58055.5539^{+0.0013}_{-0.0012}$ for the linear-$t$,
quadratic-$t$ and exponential-$t$ models, respectively. Comparing this
to the value in TK18 of $\tau = 58055.5563^{+0013}_{-0014}$, 
we consider the presence of TTVs to be validated by the new analysis,
and it is worth noting that the new reduction actually suggests the
HST transit occurred even earlier than was found in TK18.

\subsection{Summary} 

We have executed a detailed comparison of the KLB19 reduction
and that of TK18. KLB19 argue that there is in fact no evidence
for a moon in their light curve based on the absence of the moon-like dip. 
We also find that their reduction does not strongly support the presence
of the moon-like dip (although it is still formally favored using a Bayesian
model comparison), after applying the same hook correction and full Bayesian
photodynamical model selection methods used by TK18 (see Section~\ref{sub:fits}). 
We note, however, that the KLB19 reduction validates the presence of TTVs in the system,
though TTVs alone do not constitute sufficient evidence for a moon.

The question naturally arises as to why the two studies yield different results 
and which one is ultimately correct. We have argued that there are two major differences 
between the TK18 analysis and that of KLB19, and so presumably one (or both) of these is
responsible for the discrepancy. The first is the choice of systematic model
and the second is the independent reduction itself.

\subsubsection{Systematic model?}

The first major difference between TK18 and KLB19, as explored in 
Section~\ref{sub:comparisonstarmodels}, is that KLB19 use a systematic model to
correct for the long-term trend correlating flux with $X$ and $Y$
centroid position, while TK18 only decorrelate against time. TK18 found
no correlation between flux and centroid for the comparison star KIC 4760469
and in this work we have shown that amongst a broad suite of possible models,
some with and some without such correlations, models including $X$
and $Y$ correlations are consistently disfavored (see
Section~\ref{sub:comparisonstarmodels}).

Even so, this does not address whether this different choice in detrending
is ultimately responsible for the overall differing conclusions. We conclude that
the detrending choice is unlikely to be the underlying cause, since re-fitting 
the original TK18 data including centroid correlation terms still recovers the same 
exomoon signal to comparable confidence as before (see Section~\ref{sub:xyfits}).

\subsubsection{Reduction?}

With the detrending choice shown to be an unlikely explanation 
for the discrepant conclusions, we turn our attention to the reduction 
itself. There are certainly differences between the two reductions, both 
with respect to the methodologies (described in their respective papers) 
and the results. 

With regard to the methodology, the KLB19 pipeline clearly
has a track record that the TK18 reduction does not. Even so, the present observation
is unprecedented in several ways. The star is significantly fainter than previous HST targets,
the duration of the observation is far longer than typical transmission spectroscopy observations, 
and the nature of the pursued signal is fundamentally different. Therefore, it is reasonable to
ask whether the KLB19 pipeline is guaranteed to be better than the one we have developed.

We note also that the procedure for selecting an optimal aperture as described in KLB19 is potentially
problematic for the moon search. Their approach is to explore various apertures 
until they find the one which minimizes scatter with respect to the transit model.
This differs from our approach which does not assume a model.
We can only guess that given the computational expense of running a full exploration 
of parameter space with an MCMC simultaneous to the selection of an aperture, 
a static planet-only model was assumed and the scatter was minimized with respect 
to it. This approach could inadvertently incentivize the selection of an aperture for 
which the moon signal is attenuated. Nevertheless, the final aperture selected by KLB19 
is quite similar to that of TK18, the primary differences being a 13\% smaller aperture for KLB19, 
which probes slightly farther into the blue and a bit less into the red than the aperture of TK18.

We also identified anomalous behavior with orbit 22 of the HST observation as
produced by the KLB19 pipeline, which shows suspiciously low photometric scatter.
We are unable to determine the source of that anomaly, however. 

KLB19 states that the moon-like signal presented in TK18 is ``likely an artifact of the
data reduction.'' However, no faults with the original reduction pipeline were found,
nor was any step in the reduction pipeline identified as being the source of the moon-like
dip. Therefore it is perhaps more accurate to conclude (as we do here) simply that the different 
pipelines have produced different results. Of note, a recent analysis by \cite{heller:2019},
using their own independent reduction pipeline, also recovered a moon-like signal
very similar to that presented in TK18. As such, the original interpretation of the data
presented in TK18 has now been both validated and called into question in the literature.
We thus argue that the existence of the moon remains an open question and additional
observations are warranted.

To summarize 
Sections~\ref{sub:photocomp}, \ref{sub:hookcomp} and \ref{sub:centcomp},
we find that the product of the KLB19 reduction:

\begin{itemize}
\item exhibits marginally higher median intra-orbit RMS
(362.4\,ppm versus 360.6\,ppm) after correcting for the hooks,
\item has a $\simeq$900\,ppm larger flux offset at
the visit change,
\item has $\simeq$2 times larger variations in the
$y$-centroid positions,
\item has a $x$-centroid flux correlation coefficient
$\simeq3.5$ times greater, and with opposite sign to KIC 4760469,
\item exhibits a marginally higher residual RMS (370.3\,ppm vs 369.0\,ppm) after
fitting out a ``no-moon model'' and masking the claimed moon region in both
reductions
\end{itemize}

Accordingly, we argue that the KLB19 reduction is not 
obviously superior in any measurable way.

\section{Second Transiting Planet?} 
\label{sec:planetc}
\subsection{Overview}

One possible false-positive scenario for the moon-like dip that was not
discussed in TK18 was the possibility that the dip is real but
caused by a second transiting planet, not a moon. This scenario was
not investigated in the original paper because of the location of the
dip with respect to the TTV offset - indicating a strong case for the
exomoon hypothesis - as well as the inherently unlikely possibility that
a planet could have evaded detection by \kepler\ but appear in this
small segment of HST data. Nevertheless, this is certainly a valid
concern, and the probability of this scenario was not quantified in the 
original paper, so we address it here.

We express the probability that the moon-like dip was caused by a second 
(hypothetical) transiting planet, K1625c, with orbital period $P_c$, as

\begin{align}
\mathbb{P}_c = \pdf(\mathcal{T},\bar{\mathcal{D}_{\mathrm{Kep}}},\mathcal{D}_{\mathrm{HST}}|P_c),
\end{align}

where $\mathcal{T}$ is short-hand for the probability that $b<1$ (i.e. that
planet c has the correct geometry to transit), $\mathcal{D}_X$ denotes
``detected by X''. Via Bayes' theorem we can express the probability as

\begin{align}
\mathbb{P}_c &=
\pdf(\mathcal{T}|P_c)
\pdf(\bar{\mathcal{D}_{\mathrm{Kep}}}|\mathcal{T},P_c)
\pdf(\bar{\mathcal{D}_{\mathrm{HST}}}|\mathcal{T},\bar{\mathcal{D}_{\mathrm{Kep}}},P_c),\nonumber\\
\qquad&=
\pdf(\mathcal{T}|P_c)
\pdf(\bar{\mathcal{D}_{\mathrm{Kep}}}|\mathcal{T},P_c)
\pdf(\bar{\mathcal{D}_{\mathrm{HST}}}|\mathcal{T},P_c),
\label{eqn:combine}
\end{align}

where on the second-line we remove the conditional $\bar{\mathcal{D}_{\mathrm{Kep}}}$
since there is no causal dependency.

To simplify the analysis, we will assume that any other planets in the system
are coplanar with Kepler-1625b, whose low impact parameter essentially
guarantees that these planets will be transiting too. Accordingly, we
assume $\pdf(\mathcal{T}|P_c) \simeq 1$ $\forall$ $P_c \in \{\Pmin,\Pmax\}$,
where $\Pmin$ and $\Pmax$ are some yet-to-be-determined minimum/maximum
limits on the period of planet c.

This optimistic assumption of coplanarity means that we will tend to overestimate the
chance that the moon-like dip is caused by a second planet - which is the
conservative option - and reduces the overall complexity of the problem.

\subsection{Basic properties of a hypothetical K1625c}
\label{sub:basic}

The depth of the moon-like dip varies between the three different
long-term trend models adopted by TK18. In all three
cases, the radius is approximately Neptune-like, yielding
$4.90_{-0.72}^{+0.79}$\,$R_{\oplus}$ for the linear model,
$3.09_{-1.19}^{+1.71}$\,$R_{\oplus}$ for the quadratic model, and
$4.05_{-1.01}^{+0.86}$\,$R_{\oplus}$ for the exponential model.
The last value is not only the median of the three but also
represents the favored model by TK18. For this reason, we
will assume here that the hypothetical second transiting planet
has a radius of 4\,$R_{\oplus}$ in what follows.

The moon-like dip is approximately flat-bottomed, indicating that
if it were due to a transiting planet, the impact parameter is small, that is,
the planet must be non-grazing. This means the inferred radius from the depth is
a fair estimator of the true radius.

The duration of the moon-like dip varies between the models from
8.5\,hours for the linear and quadratic models to 7.8\,hours
for the exponential (using the $\tilde{T}$ transit duration definition
of \citealt{investigations:2010}). This therefore establishes that the
duration of the hypothetical K1625c must exceed 7.8\,hours. This is still
a relatively long transit duration and implies that the orbital
period is not short.

\begin{figure}
\begin{center}
\includegraphics[width=\columnwidth,angle=0,clip=true]{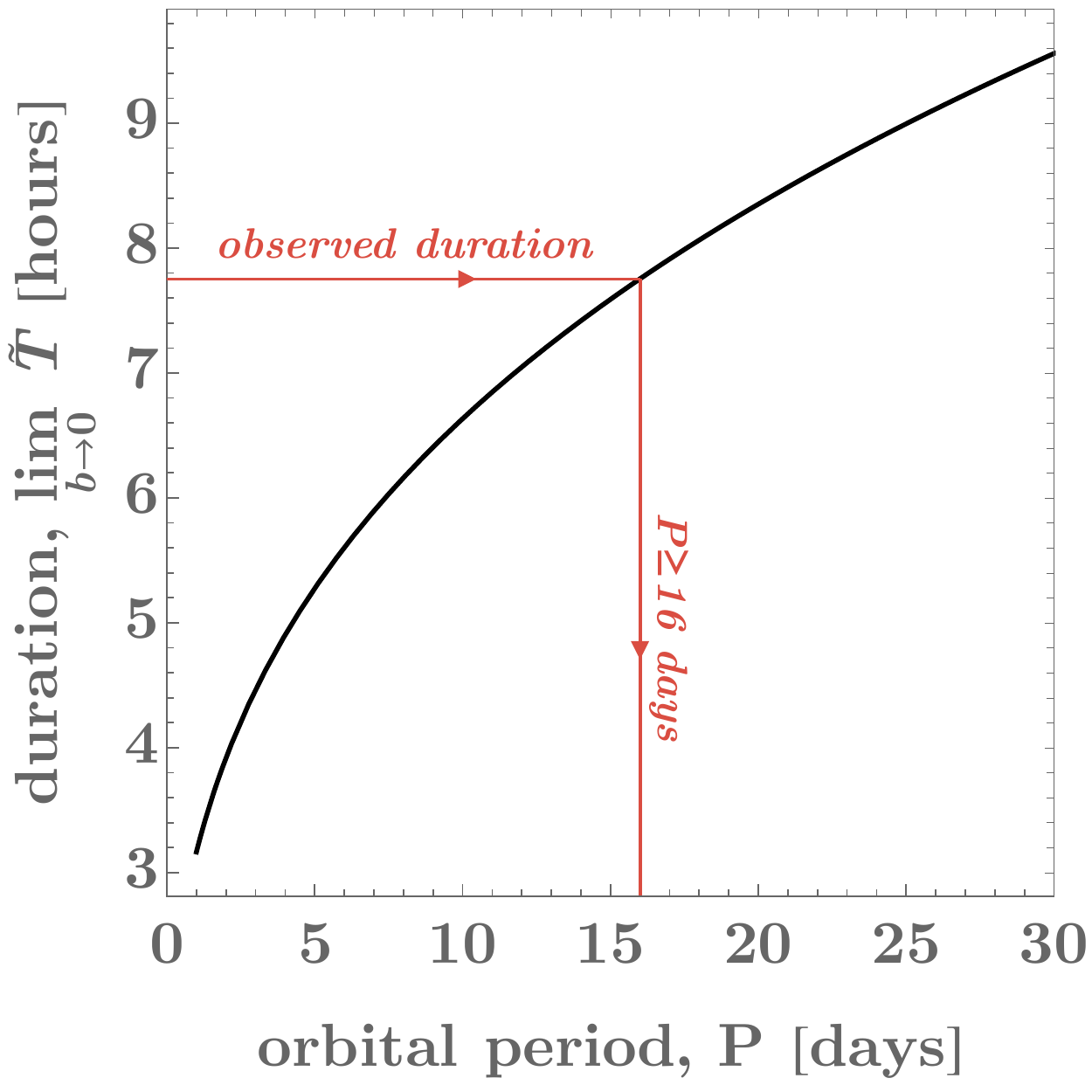}
\caption{\emph{
The moon-like dip reported by TK18 has a duration of at
least 7.8\,hours. Plotting the maximum transit duration (for
a circular orbit) as a function of period for a planet around
Kepler-1625, one can see that the period cannot be smaller than
16\,days to explain the dip.
}}
\label{fig:minP}
\end{center}
\end{figure}

For any given orbital period, the longest possible duration corresponds
to a zero impact parameter. Therefore, for any given duration, the
shortest allowed orbital period corresponds to a zero-impact parameter.
We can therefore take this duration value and convert it into a
minimum period. Assuming a circular orbit, one may solve the
\citet{investigations:2010} $\tilde{T}$ duration equation for $P$
in the limit of $b\to0$, and also transform $a/R_{\star}$ into stellar
is density, $\rho_{\star}$, using Kepler's Third Law. Since $\rho_{\star}$
is well-constrained from \gaia\ and isochrone modeling to be
$0.29_{-0.09}^{+0.13}$\,g\,cm$^{-3}$ (TK18), we can solve for the minimum 
period numerically to find $P>16$\,days, as shown in
Figure~\ref{fig:minP}.

\subsection{Probability of a missed TCE}

$\pdf(\bar{\mathcal{D}_{\mathrm{Kep}}}|\mathcal{T})$ denotes the probability
that a Neptune-sized transiting planet was undetected by the \kepler\ pipeline
- i.e. a missed threshold crossing event (TCE). There are no detected TCEs for
Kepler-1625 aside from Kepler-1625b in DR25 \citep{thompson:2017}, but
this fact alone does not provide a probability that one was missed by the
\kepler\ pipeline.

The probability of missed TCEs is most directly computed by using the
per-target detection contours for DR25 reported by \citet{burke:2017}. The
\keplerports\ software, first discussed in \citet{burke:2015}, computes
detection completeness contours for a given \kepler\ target through transit
injection and recovery tests and provides the most realistic estimate
of completeness available. The stellar parameters used by \citet{burke:2017}
are the DR25 \citet{mathur:2017} values, for which Kepler-1625 is reported
as a $1.79_{-0.49}^{+0.26}$\,$R_{\odot}$ - which is approximately the same
as the \gaia-based value found by TK18 of
$1.73_{-0.22}^{+0.24}$\,$R_{\odot}$. This therefore demonstrates that the
\keplerports\ detection contours for a given planetary size do not require
any significant update since the minor revision of TK18.

\begin{figure}
\begin{center}
\includegraphics[width=\columnwidth,angle=0,clip=true]{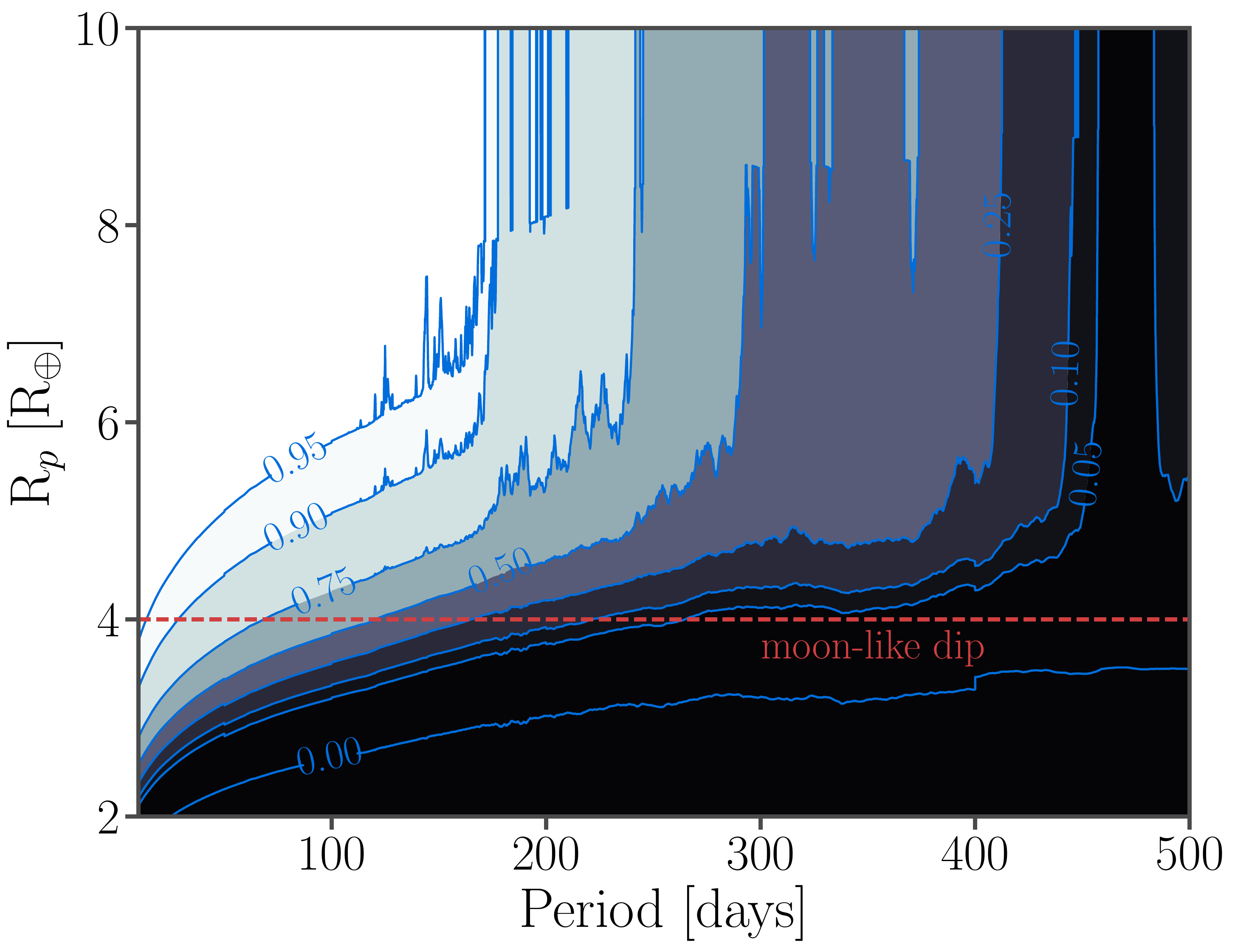}
\caption{\emph{
Detection completeness contours plot generated by \keplerports\
\citep{burke:2015,burke:2017} for the target Kepler-1625.
}}
\label{fig:burke}
\end{center}
\end{figure}

After running \keplerports\ on our target (see Figure~\ref{fig:burke}), we
extracted a slice along the radius axis of 4\,$R_{\oplus}$, corresponding
to the dip seen in the HST data by TK18. This is shown in the left-most
panel of Figure~\ref{fig:planetc}.

\keplerports\ natively computes completeness only out to 500\,days, and indeed
by this point the probability of missing a 4\,$R_{\oplus}$ exceeds 99\% and is
effectively unity - meaning there is little point in extending past this
period.

\subsection{Probability of K1625c transiting in the HST window}

If the moon-like dip were due to another planet, then within the HST window
of $W=38.8$\,hours we would have observed a single transit of our hypothetical planet K1625c. 
The HST photometry is approximately four times superior to that of \kepler\
and thus HST is effectively complete to a Neptune-sized transit of the observed
duration. $\pdf(\mathcal{D}_{\mathrm{HST}}|\mathcal{T})$ then simply
reduces to the probability that the planet will have the correct phasing to
transit within the 38.8\,hour observing window.

Consider the possibility that K1625c has an orbital period of $100$\,years.
The chance of seeing this world transit in a fixed window of observations
is clearly going to be very low. Indeed, the chance of seeing a planet with
period $P$ transit at least once in a window is $\propto 1/P$. This is known
as the window effect and is described in detail in \citet{kipping:2018},
who shows that

\begin{equation}
\begin{split}
&\pdf(n=1|P_c,W,\mathcal{T}) = \\
&\pdf(n\geq1|P_c,W,\mathcal{T}) (1-\pdf(n\geq2|P_c,W,\mathcal{T})),
\label{eqn:single}
\end{split}
\end{equation}

where $n$ is the number of transits observed in the window of duration $W$
and the components probabilities are

\begin{equation}
\pdf(n\geq1|P_c,W,\mathcal{T}) =
\begin{cases}
1  & \text{if } P_c \leq W,\\
\frac{W}{P_c} & \text{if } P_c > W,
\end{cases}
\end{equation}

and

\begin{equation}
\pdf(n\geq2|P_c,W,\mathcal{T}) =
\begin{cases}
1  & \text{if } P_c \leq \tfrac{W}{2},\\
\frac{W-P_c}{P_c} & \text{if } \tfrac{W}{2} < P_c \leq W ,\\
0 & \text{if } P_c > W .
\end{cases}
\end{equation}

\citet{kipping:2018} shows how a lower limit on the period can be derived
from the relative phase of the transit within the window but in our case
a far more constraining lower limit on the period comes from the duration
argument earlier in Section~\ref{sub:basic}. Imposing this as a hard limit
simplifies Equation~(\ref{eqn:single}) to

\begin{align}
\pdf(n=1|P_c,W,\mathcal{T}) = \frac{W}{P}.
\end{align}

This is shown in the middle panel of Figure~\ref{fig:planetc}.
And finally we may write that $\pdf(\mathcal{D}_{\mathrm{HST}}|\mathcal{T}) =
\pdf(n=1|P_c,W,\mathcal{T})$ since we treat HST as effectively complete
to Neptune-sized transits.

\subsection{Combining the constraints}   

\begin{figure*}
\begin{center}
\includegraphics[width=2\columnwidth,angle=0,clip=true]{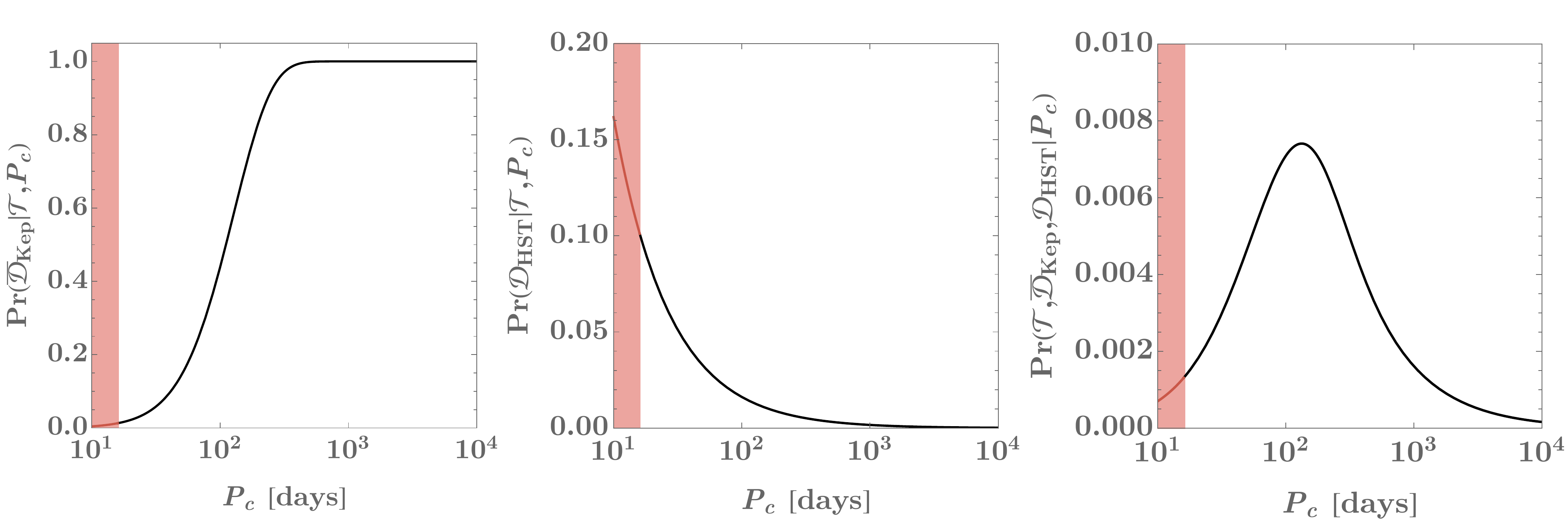}
\caption{\emph{
Probability that a Neptune-sized transiting planet evaded detection by \kepler\
(left), was seen to transit in the HST window of TK18 (middle),
and the probability of both of these statements being true - as a function of
the planet's orbital period.
}}
\label{fig:planetc}
\end{center}
\end{figure*}

The final step is to combine the probabilities from above using
Equation~(\ref{eqn:combine}), which is shown in the right-most panel
of Figure~\ref{fig:planetc}. The probability peaks at $P_c=133.3$\,days
with $\mathbb{P}_c = 0.74$\%, and decreases monotonically either side.
Given the presence of TTVs in the system (TK18), the most plausible
planet-scenario to explain both the dip and the TTVs would be an interior
transiting planet close to a mean motion resonance (e.g. 2:1 would lead
to $P_c \simeq 144$\,days).

The probability computed above suggests the existence of another
transiting planet causing the moon-like dip is quite low, which might 
lend additional credence to the exomoon hypothesis. At the same time,
this probability should be weighed against the probability of observing such 
a large exomoon. This comparison unfortunately eludes us for the time being, 
as there are so far no other verified exomoons in the literature, and
the occurrence rate of such an unanticipated object cannot be quantified
at this time.

\section{Stellar Activity} 
\label{sec:stellar}
\subsection{Rotation}
\label{sub:rotation}

There is no known rotation period for Kepler-1625 at the time of writing.
The star is included within the autocorrelation function (ACF) catalog of
\citet{mcquillan:2014}, but no clear rotation period was found in that work.

\begin{figure*}
\begin{center}
\includegraphics[width=2\columnwidth,angle=0,clip=true]{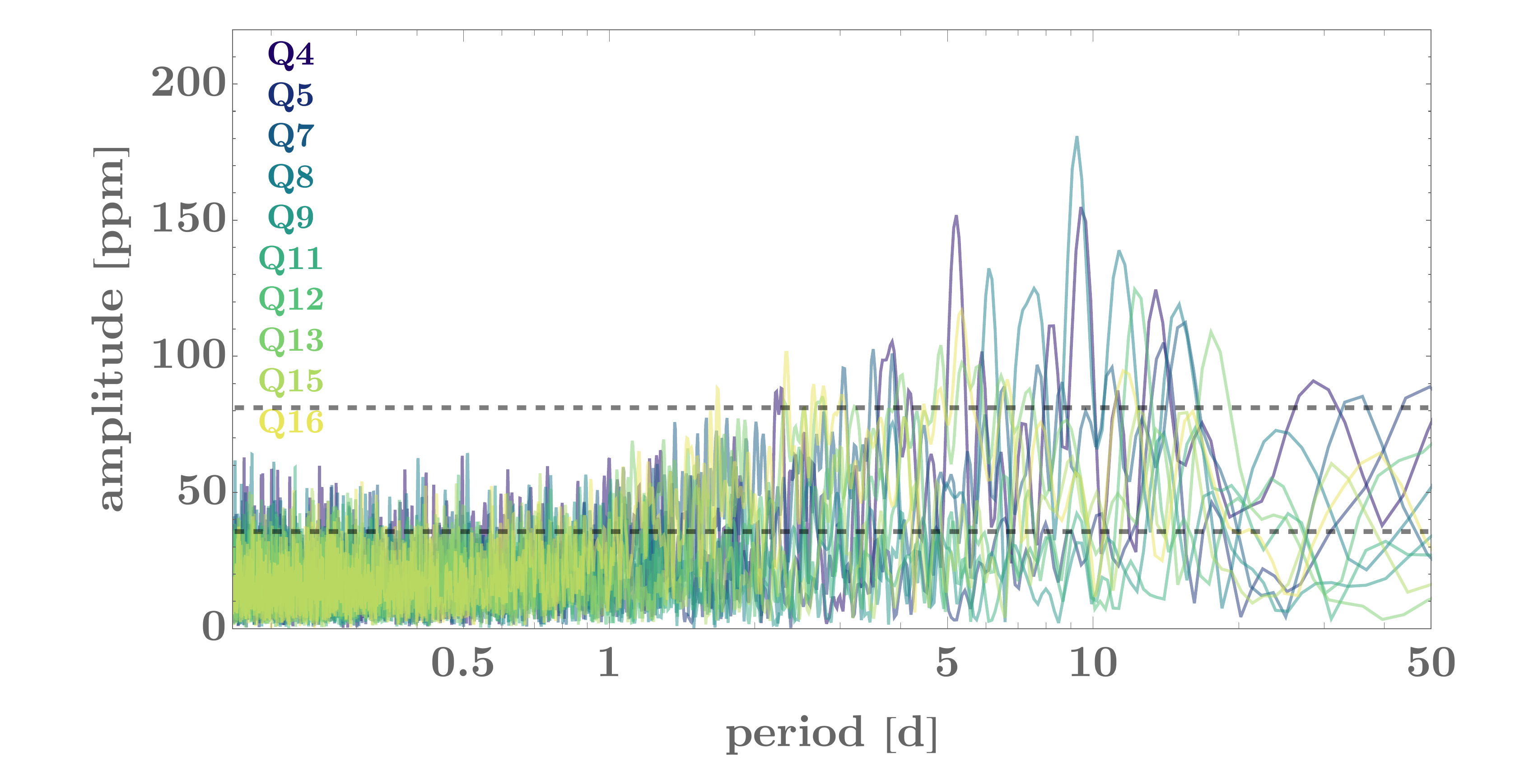}
\caption{\emph{
LS periodogram for the PDC data of Kepler-1625 for ten
quarters. Activity appears lower than 200ppm for periods $<50$ days. Dashed lines represent the $p$-values of 0.05 
for the most active and most quiescent quarters based on bootstrapping. Activity above these limits may be considered real, but a clear rotation period across all quarters is not detected.
}}
\label{fig:periodogram}
\end{center}
\end{figure*}

We attempted to search for the rotation period using a Lomb-Scargle (LS) periodogram, 
applying the algorithm to each \kepler\ quarter (PDC data) independently.
Since each quarter is treated independently, and each quarter has a duration of
$\simeq 90$\:days, it is not possible to detect periods longer than approximately
half this value. The results out to 50\:days are therefore shown in
Figure~\ref{fig:periodogram}.

Consistent with the analysis of \citet{mcquillan:2014}, we are unable to
identify any clear rotation period from the LS periodogram.
We find that the maximum amplitude of a periodic signal $<50$\:days must
be less than 200\,ppm, significantly lower than the amplitude of the
moon-like dip reported by TK18.

We also attempted to recover a rotation period using Gaussian process regression. We used the \celerite\ 
software package \citep{celerite} to model the light curve, the kernel function consisting
of a mixture of two simple harmonic oscillators with periods separated by a factor of two 
\footnote{see \url{https://exoplanet.readthedocs.io/en/latest/tutorials/stellar-variability/}}. 
Exploring the posterior PDF of the star's rotation period using \texttt{PyMC3} \citep{pyMC3} we infer a rotation 
period of $12.9^{+0.7}_{-0.6}$ days. However, for this period the natural log of the $Q$ factor, 
or damping ratio, was $-3.1 \pm 0.3$. This means that the light curve, when modeled as a damped 
harmonic oscillator, is overdamped ($Q \sim 0.05$), indicating that the stellar brightness variations 
are incoherent, which suggests that the star spot lifetimes are shorter than the rotation period 
of the star on average. This, combined with the inability of both the LS periodogram and the ACF to recover 
a reliable period, implies that the signal is aperiodic and non-sinusoidal. Taken together with the low RMS of 
the light curve ($\sim 200$\:ppm), this indicates that Kepler-1625 is an inactive star and that there is 
little evidence of short-timescale (sub-hour) variability that could mimic the ingress of a moon.

Although not a direct measure of the rotation period, the $v \sin i$ can
provide some useful information on rotation too. We obtained two Keck
\textit{HIRES} spectra without iodine in October and November 2018 to attempt
to measure the velocity broadening. Using the \specmatch\ pipeline described in
\citet{specmatch}, we obtain a marginal detection of $v \sin i =
(1.9\pm1.0)$\,km/s. Combining this with the isochrone posteriors from
TK18 yields a minimum rotation period of $45_{-15}^{+44}$\,days
(a minimum because we don't know $\sin i$). It is therefore probable that
the rotation period falls within the 50\,day range that \kepler\ is sensitive
to, but that the amplitude of rotational modulations is simply too small
to reliably recover.

\subsection{Activity-induced dips}

Stellar activity can produce complex morphological signatures in photometric
time series \citep{soap:2014}. Although the photometric periodic behavior
of Kepler-1625 appears limited to $<200$\,ppm (see Section~\ref{sub:rotation}) 
-- too small to induce an effect comparable to the moon-like transit reported by
TK18 -- shorter, non-periodic variations deserve our attention.

The moon-like dip is characterized by a transit depth of $\simeq$500\,ppm in
the integrated light (white) bandpass of WFC3 (TK18). As a near-infrared
instrument, stellar activity is generally expected to be suppressed by WFC3
versus an optical bandpass like that of \kepler. To estimate the magnitude of
this effect, we took the isochrone posterior chains for Kepler-1625
(TK18) and extracted the median effective temperature of the star,
$T_{\mathrm{eff}}=5563$\,K. We then assume spots on the surface with a
temperature approximately 2000\,K cooler than the photosphere, typical of
sunspots. Integrating a Planck function multiplied by the bandpass response
function for \kepler\ and WFC3 reveals that spots would appear 1.3 times
larger in amplitude, as viewed by \kepler\ than WFC3. Accordingly, if the
moon-like dip is due to spots, then one should expect to see frequent dips
of amplitudes of $\simeq650$\,ppm (500 $\times$ 1.3) in the eleven quarters of
\kepler\ data for Kepler-1625.

\begin{figure*}
\begin{center}
\includegraphics[width=2\columnwidth,angle=0,clip=true]{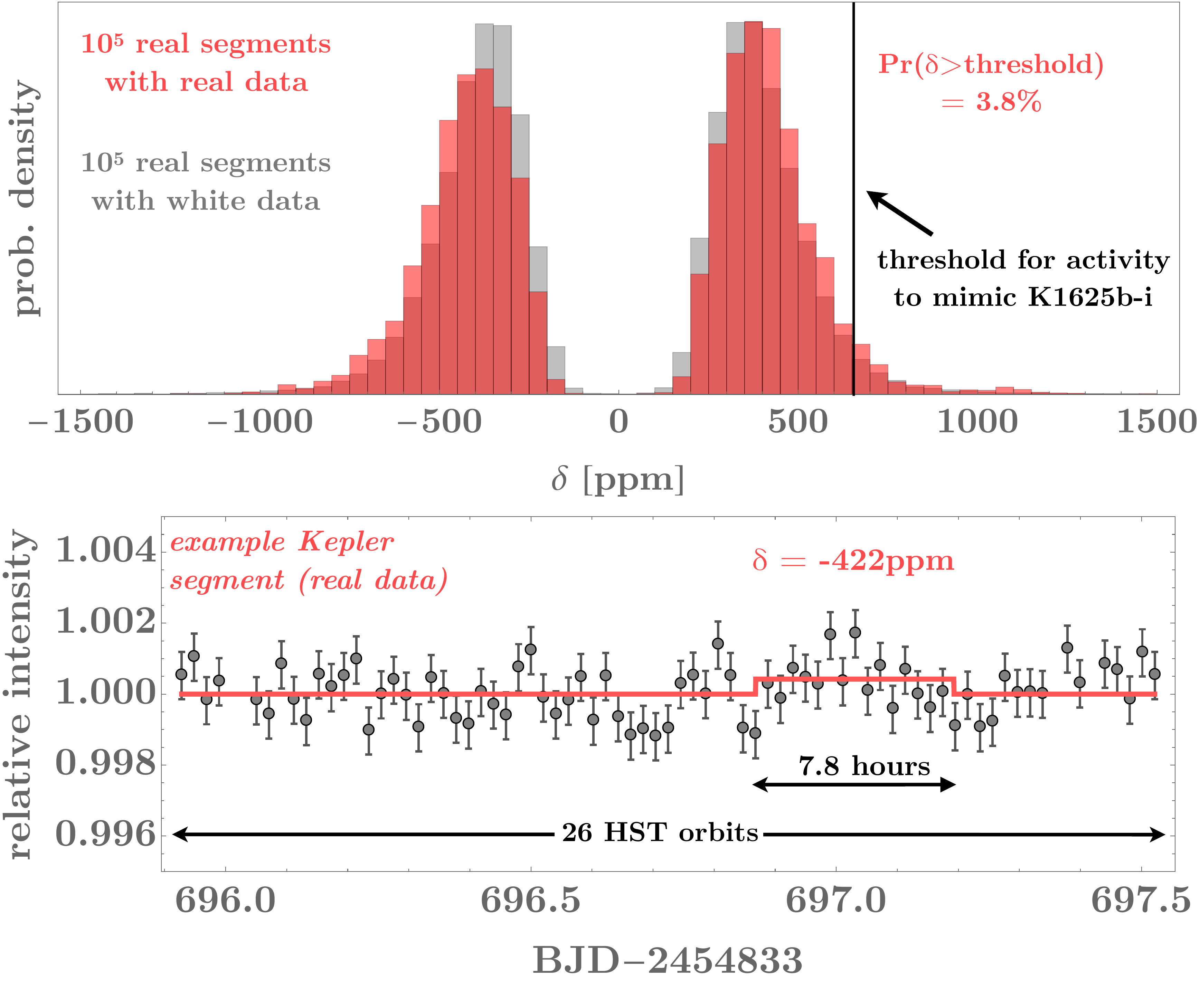}
\caption{\emph{
Bottom: Example of a random \kepler\ segment of Kepler-1625 with a duration
equal to that of the TK18 HST observing window. Regressing the best
fitting box with a duration of 7.8\,hours (same as the moon-like dip) finds
an inverted transit in this case of $422$\,ppm depth.
Top: Repeating this exercise on $10^5$ random segments, we obtain a nearly
symmetric distribution of best-fitting box amplitudes (red histogram). For
comparison, we repeated the simulations assuming pure Gaussian noise only
(gray histogram), which is nearly identical.
}}
\label{fig:boxes}
\end{center}
\end{figure*}

To test this hypothesis, we extracted a random segment from a random quarter
of the available \kepler\ PDC time series, with the segment duration being
equal to 26 HST orbits (which was the time window observed by TK18).
Each quarter was first detrended using a median filter of window size equal
to 5 times the minimum duration of the moon-like dip, approximately
7.8\,hours (TK18). We then performed a blind search for the best-fitting
box-like transit within this segment, forcing the box to have a duration equal
to that of the TK18 moon-like dip. The central time and depth were
optimized for in a least squares sense. An example of this is shown in lower panel of
Figure~\ref{fig:boxes}.

The best-fitting box was saved and then a new random segment was picked,
repeating $10^5$ times. A histogram of the best-fitting depths is shown in the
upper panel of Figure~\ref{fig:boxes}. Because this is the best-fitting depth
within a segment, these depths always deviate from zero since the regression
routine is allowed to try many different possible central times. There is
symmetry about zero, with just as many inverted transits as positive transits
being recovered. We find that 3.8\% of the experiments run on the \kepler\
data are able to produce a best-fitting transit of depth exceeding 650\,ppm
(and 3.5\% produce depths $<-650$\,ppm). Naively, one might interpret this as
indicating that the moon-like dip reported by TK18 is only 2.1\,$\sigma$
significant (3.8\%). However, these simulations were conducted for the 0.95\,m
\kepler\ telescope data, and not the 2.4\,m HST data set in which the dip is
actually observed. To interpret this 3.8\% number, one must consider the plausible 
origin of these spurious (i.e. moon-mimicking) events.

If indeed the signals are spurious, there are two possible causes for these
random quasi-dips. Either 1) time-correlated noise structure caused by intrinsic
stellar activity is able to produce $>650$\,ppm dips, or 2) the noise is not
significantly correlated (i.e. white noise) but the noise budget of the
\kepler\ photometry is sufficiently large that the best-fitting boxes can
infrequently exceed 650\,ppm.

If the former were true, then the dip observed by HST could be explained as
one of these 3.8\% instances of an activity-driven false-positive. If the
latter were true, then one could expect it to be highly improbable
for the HST moon-like dip to be a product of Gaussian noise, as the
measurement uncertainties are 3.8 times smaller than that of \kepler.

Clearly this is an important distinction. To distinguish between them,
we can set up another experiment where we repeat our previously described
Monte Carlo experiments except we replace the real \kepler\ photometric
fluxes with artificial fluxes computed assuming pure Gaussian (white) noise.
The artificial data are drawn from a normal distribution with a mean of unity
and standard deviation equal to the standard deviation of a randomly
picked real \kepler\ segment.

After drawing $10^5$ segments and replacing the photometric fluxes with
white noise, we make a histogram of the best-fitting box depths as before
and find a very similar distribution, as shown in Figure~\ref{fig:boxes}.
The 650\,ppm threshold is exceeded in a similar number of trials, 2.4\%.
We interpret the similarity between these two distributions as evidence
favoring the hypothesis that the spurious, moon-mimicking detections are 
simply a product of Gaussian-like noise controlled by photon counting statistics, 
rather than being due to intrinsic stellar activity. Since the HST data is 
much more precise, the probability of a white noise driven box is far smaller, 
and is in fact accurately accounted for in our evidence calculations since 
we assumed a normal likelihood function in TK18. We therefore conclude that 
there is no evidence from the \kepler\ analysis that activity is a plausible
explanation for the moon-like reported by TK18 in the HST data.

\section{Follow-Up} 
\label{sec:followup}
\subsection{Photometric follow-up}

The best way to confirm the presence of the exomoon candidate would be 
to see it transit again. To this end we have explored various avenues 
to observe future transits of Kepler-1625b. Unfortunately, this is a 
very challenging target for transit observations because of its faintness ($K_P=15.756$) 
and the very long duration of the planet's transit ($\sim$ 19\,hours). These 
challenges are exacerbated by the fact that the exact location of the 
exomoon cannot be known ahead of time for any given transit; as we project 
into the future, our predictions are naturally degraded as the uncertainties 
in our posterior samples propagate. A wide range of times before planetary 
ingress and after planetary egress must therefore be monitored to cover as many
geometries as possible. 

These limitations generally restrict any efforts to detect the 
exomoon transit to space-based telescopes. However, targeted observations 
of this sort clearly require considerable dedicated resources to the exclusion 
of other priorities. While \textit{Spitzer} may be a suitable alternative to
observing with HST, the former can only observe $\sim$ 35\% of the sky at any given
time due to pointing restrictions, and cannot observe the May 2019 transit as 
the target falls within the zone of avoidance. Future observations carried out 
with a survey (non-targeted) spacecraft could potentially bear fruit, though we 
note that the Transiting Exoplanet Survey Satellite (TESS) observation of the 
\kepler\ field will occur in July 2019, missing the May 2019 transit of Kepler-1625b. 

On the other hand, transit timings could potentially be measured from the ground more 
easily, and continued monitoring of the TTVs and (to the extent possible) 
transit duration variations (TDVs) would be valuable. A single instrument may 
be able to monitor in its entirety either planetary ingress or egress, but likely 
not both, due to the time separation of these two events. Of course, this requires the 
target to be up at night long enough for the observation to be made, and the telescope 
must be located at a longitude where the event can be observed in its entirety without 
sunset or sunrise encroaching. Latitude is also a consideration; while northern latitudes 
place the target above the horizon for longer durations, they also experience a greater 
range of night lengths.

Radial velocity (RV) measurements of the system may also provide additional evidence for or 
against the moon hypothesis. On the one hand, RVs could potentially yield evidence for a 
second planet in the system, in which case the observed TTVs might be attributable to that 
planet and the case for the moon would be weakened. On the other hand, if an additional 
massive planet can be ruled out, or strong constraints can be placed on the mass and 
location of an undetected planet, the moon could emerge as a stronger candidate insomuch as
an alternative mechanism for the timing variations is weakened or removed altogether. 

\begin{figure}
    \centering
    \includegraphics[width=8cm]{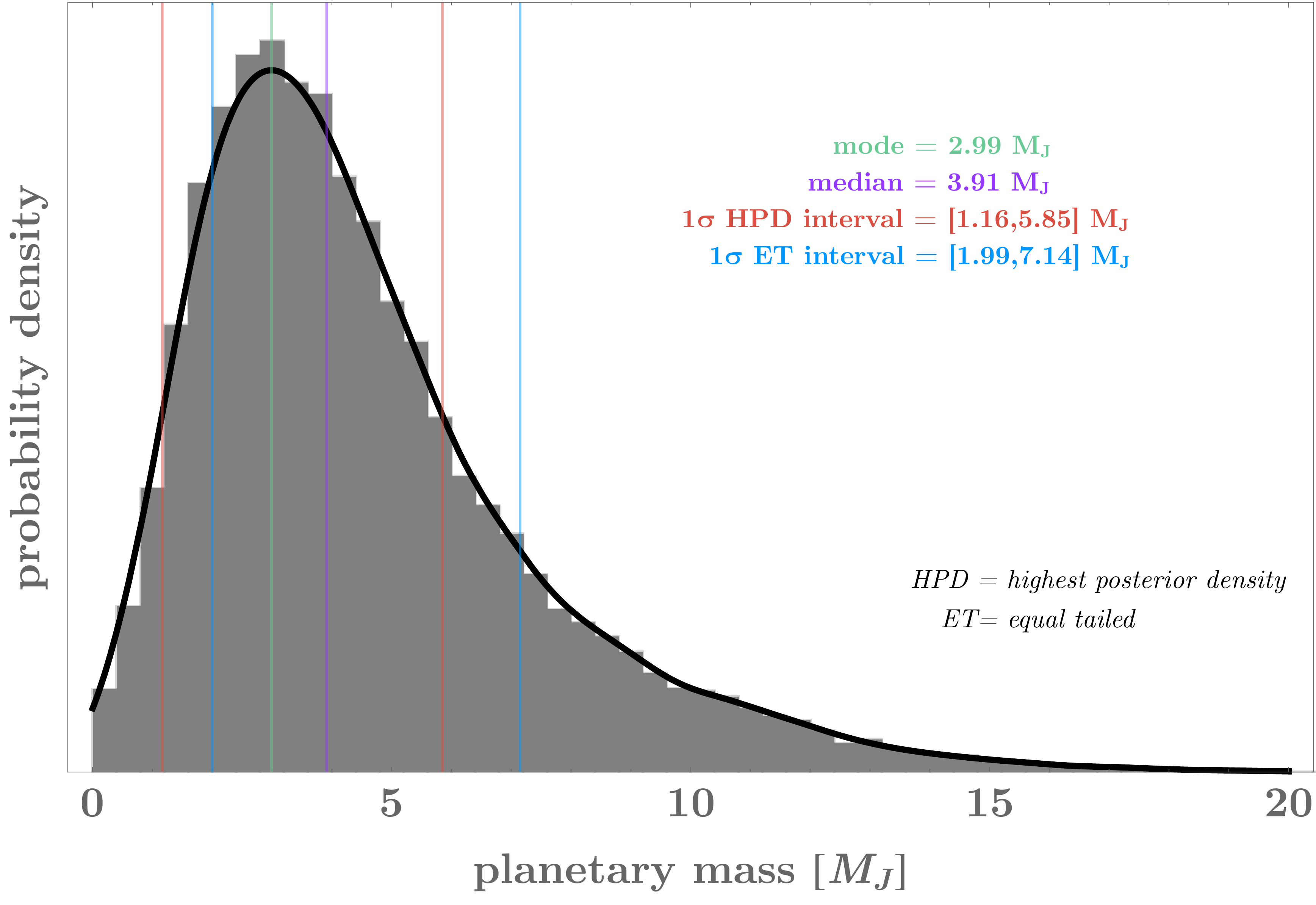}
    \caption{Combined mass posterior distribution for Kepler-1625b.}
    \label{fig:k1625_mass_estimate}
\end{figure}

Of course RV measurements should also provide a reliable measurement of the target planet's mass. 
If Kepler-1625b is revealed to be significantly less massive than anticipated, this 
could also weaken the moon case, as it would be more difficult to support such a large moon.
Conversely, a mass measurement consistent with the inferred mass presented in TK18 could
lend additional credence to the moon hypothesis. Figure~\ref{fig:k1625_mass_estimate} presents 
our best estimate for the planet's mass which may be tested by the acquisition of RVs.

An additional complication for photometric confirmation through transit observations 
arises from the exomoon candidate's inferred large inclination with respect to
the planet's orbital plane (TK18 found $i_S = 42^{+15}_{-18}$, $49^{+21}_{-22}$, and
$43^{+15}_{-19}$ degrees for the linear, quadratic, and exponential detrendings, respectively). 
This has the effect of sending the moon high above or below the disk of the star for a 
significant fraction of its orbit, precluding the possibility of a transit when the moon is 
in these positions. Coupled with the uncertainty in the moon's true anomaly, this means that
for any given transit observation there is no guarantee of seeing the moon transit at 
all. Thus, a null detection of the moon for any given epoch cannot be interpreted 
as definitive evidence that the moon does not exist. Only with many repeated 
observations, all lacking evidence for the moon, could the moon truly be ruled 
out to high confidence \citep{Martin:2019}. Clearly this has a multiplicative impact on the 
telescope requirements, and naturally leads to the conclusion that follow-up transit 
observations are only worthwhile to the extent that they are not excessively 
expensive.

\begin{figure*}
\begin{center}
\includegraphics[width=2\columnwidth,angle=0,clip=true]{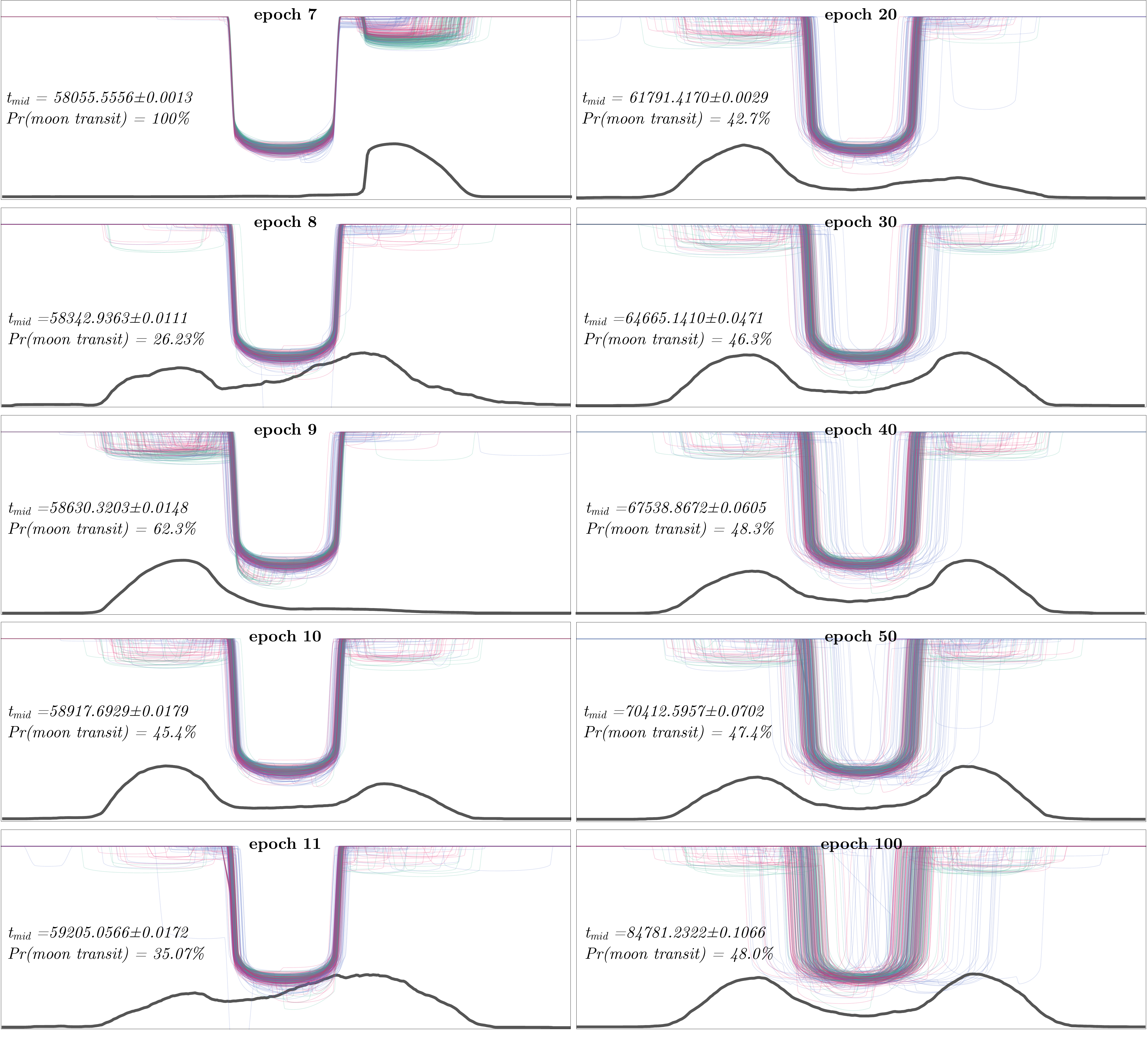}
\caption{\emph{
Projections of the light curve of Kepler-1625b and its candidate moon
into the future, using the posterior samples of TK18. We show 100 light curves
from each trend model (green = linear, blue = quadratic and pink =
exponential). Uncertainties can be seen to visually grow over time. The
gray distributions reflect the most likely location for a moon transit.
}}
\label{fig:projections}
\end{center}
\end{figure*}

Using the posterior samples from \citet{TK18}, it is possible to predict
the morphology of the combined planet and moon system for future transit
events, though as noted these predictions naturally deteriorate with every epoch.
We elected to consider ten epochs, including the original observed epoch for 
comparison, and calculate 1000 projections of the transit light curve for 
Kepler-1625b. For this purpose, we used model
$\mathcal{M}$ and repeated for each of the three trend models used by
\citet{TK18}. The light curves can be viewed in
Figure~\ref{fig:projections}.

For each epoch, we find the time of transit minimum for the planet
component only and use these times to compute a median mid-transit
time and an associated standard deviation, which is quoted in
the panels of Figure~\ref{fig:projections}. We also consider the moon
component in isolation and count how often the moon presents any
deviation in flux away from unity - the probability that the moon
will transit at all in each epoch. These probabilities are again added to
the panels of Figure~\ref{fig:projections}. Finally, we use the moon
component only to compute a probability distribution for the most
likely location one should expect to observe the exomoon (assuming it
transits at all). These curves are shown in gray in
Figure~\ref{fig:projections}.

As expected, we find that the uncertainty in the time of transit
grows as we project further into the future. Regressing a power-law
to the uncertainties, we find that uncertainty grows as $t^{2/13}$
to an excellent approximation. The probability of a moon transiting
oscillates for the first few epochs but then tends towards slightly
less than 50\%, which is broadly consistent with the findings 
of \citet{Martin:2019}. Observers can therefore treat the chances of seeing
the moon in a given future transit as approximately 1 in 2.

We also highlight that epoch 9 (May 2019) appears to be the most
favorable for follow-up. It has the highest probability of seeing a moon
transit out of any future epoch and a clean prediction for the location of
said transit (before planetary ingress). A proposed observation of this transit
with HST was not awarded, and as pointed out, there are no other viable space-based
options for this event.

In light of these challenges, photometric confirmation of the exomoon 
candidate may remain elusive for some time, until repeated observations
may be performed at relatively low cost. Of course, if the moon is real,
eventual confirmation is probably inevitable, but in the near term it will 
likely remain merely a candidate.

\section{Conclusions}  
\label{sec:conclusions}
In this work we have examined a number of alternative hypotheses
put forth by the community to explain the two critical pieces of the 
exomoon case for Kepler-1625b, namely, the presence of significant 
TTVs, and a sustained flux reduction in the HST light curve following 
planetary egress. We have explored various additional detrending models, 
employing more degrees of freedom, and found that while some of these 
approaches are able to attenuate the purported moon signal, this is 
to be expected given their flexibility. That is, from the standpoint
of their Bayesian evidences, more flexible detrending models paired 
with planet-only transit models are in some cases indistinguishable from 
simpler detrending models combined with system models that include a moon.
While we cannot rule out the presence of unprecedented systematic effects,
we also see no evidence for them, and therefore the adoption of more flexible
detrending models that attenuate the moon signal are not particularly well 
motivated.

We have investigated the differences between the light curve presented in 
TK18 and a new reduction from \citet{kreidberg}, and find that
while the source of the discrepancy is not readily identifiable,
our light curve displays effectively identical noise properties,
and therefore, the KLB19 light curve is not demonstrably superior. We
also highlight once again the work of \citet{heller:2019} which, through
their own independent reduction and analysis, also found evidence of a 
moon-like dip following planetary egress.

In terms of a possible additional transiting planet in the system, we 
have calculated the probability that such a planet could have gone 
undetected in the \kepler\ data and transit in the short time window 
of the HST observation and find the maximum probability of this scenario 
to be $< 0.75\%$. 

To determine whether the dip in brightness measured with HST could be due to
stellar activity, we have attempted to measure a rotation period for the star in
the \kepler\ data using a variety of standard methods and are unable to recover
it, indicating that the star exhibits negligible periodic variability. We have also
searched for photometric dips that might be associated with (non-periodic) star spot 
crossings and find that such dips, while possible to find in the \kepler\ data, 
are consistent with Gaussian noise. 

Finally, we discussed the outlook for confirming the presence of the exomoon 
using space-based transit monitoring, radial velocity observations, and ground-based
measurement of transit timings. We find that the system poses a number of substantial challenges
to observational confirmation in the near-term, and conclude that while modest ground-based
observations may be worthwhile for 1) constraining the mass of the planet, 2) quantifying the probability of 
an unseen perturber in the system, and 3) measuring TTVs, additional targeted observations from 
space likely fail a reasonable cost-benefit analysis. Confirming or refuting the moon to high 
confidence may therefore require many years and the advent of additional space-based time-domain survey 
data that can be acquired at minimal cost.

\section{Acknowledgements}
We would like to thank Laura Kreidberg for useful discussions
and providing data products in advance of her paper, which allowed 
us to investigate in greater detail the source and degree of the 
discrepancies between her findings and ours. We also thank Erik Petigura 
for performing the $v \sin i$ measurement. We thank the anonymous reviewer
for thorough comments which strengthened this paper. Finally, we wish to thank 
past and present HST and \textit{Kepler} scientists and engineers, mission support
personnel, and the crews of STS-31, 61, 82, 103, 109, and 125, who through their
dedication have been jointly responsible for making this work possible.

Analysis was carried out in part on the NASA Supercomputer PLEIADES
(Grant \#HEC-SMD-17-1386).
AT is supported through the NSF Graduate Research Fellowship (DGE-1644869). 
DK is supported by the Alfed P. Sloan Foundation Fellowship.
This work is based in part on observations made with the NASA/ESA Hubble Space Telescope,
obtained at the Space Telescope Science Institute, which is operated by the
Association of Universities for Research in Astronomy, Inc., under NASA contract
NAS 5-26555. These observations are associated with program \#GO-15149.
Support for program \#GO-15149 was provided by NASA through a grant from the
Space Telescope Science Institute, which is operated by the Association of
Universities for Research in Astronomy, Inc., under NASA contract NAS 5-26555.”
This paper includes data collected by the \kepler\ mission. Funding for the 
\kepler\ mission is provided by the NASA Science Mission directorate.



\end{document}